\documentclass[twocolumn]{aastex62}

\usepackage{graphicx}	
\usepackage{amsmath}	
\usepackage{amssymb}	

\def\equationautorefname~#1\null{equation~(#1)}

\newcommand{\mbh}{\langle m_{\mathrm{\bullet}}\rangle}
\newcommand{\mstar}{\langle m_{\star}\rangle}
\newcommand{\Mcl}{M_{\mathrm{cl}}}
\newcommand{\Mbh}{M_{\mathrm{\bullet}}}
\newcommand{\rev}[1]{\textcolor{black}{#1}}

\newcommand{\Mh}{M_{\mathrm{h}}}
\newcommand{\Msun}{\,{\rm M}_{\odot}}

\newcommand{\Mstar}{M_{\star}}

\newcommand{\feh}{\mathrm{[Fe/H]}}
\newcommand{\logz}{\log_{10}Z/Z_{\odot}}
\newcommand{\sigbh}{\sigma_{\bullet}}
\newcommand{\sigstar}{\sigma_{\star}}
\newcommand{\aej}{a_{\rm ej}}
\newcommand{\agw}{a_{\rm GW}}
\newcommand{\tthree}{\tau_{\mathrm{3bb}}}
\newcommand{\tsb}{\tau_{1-2}}
\newcommand{\tdis}{\tau_{\mathrm{evap}}}
\newcommand{\ttid}{\tau_{\mathrm{tid}}}
\newcommand{\tgw}{\tau_{\mathrm{GW}}}
\newcommand{\ttot}{\tau_{\mathrm{tot}}}
\newcommand{\tdf}{\tau_{\mathrm{df}}}
\newcommand{\kms}{\mathrm{km\,s^{-1}}}
\newcommand{\acrit}{a_{\mathrm{crit}}}
\newcommand{\ah}{a_{\mathrm{hard}}}
\newcommand{\tbhb}{\tau_{\mathrm{BHB}}}

\newcommand{\tcrit}{\tau_{\mathrm{crit}}}
\newcommand{\dt}{\Delta t_{z=0}}
\newcommand{\rc}{r_{\rm c}}
\newcommand{\rh}{r_{\rm h}}
\newcommand{\rhostarc}{\rho_{\rm \star,c}}
\newcommand{\rhobhc}{\rho_{\bullet, \rm c}}
\newcommand{\tdes}{\tau_{\rm dis}}
\newcommand{\thub}{t_{\rm H}}

\newcommand{\gpcyr}{\mathrm{Gpc}^{-3}\,\mathrm{yr}^{-1}}
\newcommand{\Mpcc}{\mathrm{Mpc}^{-3}}

\graphicspath{{./}{figures/}}

\submitjournal{ApJ}

\shorttitle{BHBs in GCs across time}
\shortauthors{Choksi et al}

\begin{document}

\title{\large \bf The star clusters that make black hole binaries across cosmic time}

\correspondingauthor{N. Choksi}
\email{nchoksi@berkeley.edu}

\author{Nick Choksi \href{https://orcid.org/0000-0003-0690-1056}{\includegraphics[scale=0.4]{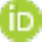}}}
\affiliation{University of California, Berkeley, Department of Astronomy, Campbell Hall, Berkeley, CA 94704, USA}

\author{Marta Volonteri \href{https://orcid.org/0000-0002-3216-1322}{\includegraphics[scale=0.4]{orcid.pdf}}}
\affiliation{Institut d'Astrophysique de Paris, UMR 7095 CNRS, Universit\'{e}
Pierre et Marie Curie, 98bis Blvd Arago, 75014 Paris,
France}

\author{Monica Colpi \href{https://orcid.org/0000-0002-3370-6152}{\includegraphics[scale=0.4]{orcid.pdf}}}
\affiliation{Department of Physics, University of Milano Bicocca and Istituto Nazionale di Fisica Nucleare-Sezione di Milano Bicocca, Piazza della Scienza 3, I-20126 Milano, Italy}

\author{Oleg Y. Gnedin \href{https://orcid.org/0000-0001-9852-9954}{\includegraphics[scale=0.4]{orcid.pdf}}}
\affiliation{Department of Astronomy, University of Michigan, Ann Arbor, MI, 48109, USA}

\author{Hui Li \href{https://orcid.org/0000-0002-1253-2763}{\includegraphics[scale=0.4]{orcid.pdf}}}
\affiliation{Department of Physics, Kavli Institute for Astrophysics and Space Research, MIT, Cambridge, MA 02139, USA}



\begin{abstract}
We explore the properties of dense star clusters that are likely to be the nurseries of stellar black holes pairing in close binaries. We combine a cosmological model of globular cluster formation with analytic prescriptions for the dynamical assembly and evolution of black hole binaries to constrain which types of clusters are most likely to form binaries tight enough to coalesce within a Hubble time. We find that black hole binaries which are ejected and later merge ex-situ form in clusters of a characteristic mass $\Mcl \sim 10^{5.3}\Msun$, whereas binaries which merge in-situ form in more massive clusters, $\Mcl \sim 10^{5.7}\Msun$. The clusters which dominate the production of black hole binaries are similar in age and metallicity to the entire population. Finally, we estimate an approximate cosmic black hole merger rate of dynamically assembled binaries using the mean black hole mass for each cluster given its metallicity. We find an intrinsic rate of $\sim 6\,\gpcyr$ at $z=0$, a weakly increasing merger rate out to $z=1.5$, and then a decrease out to $z=4$. Our results can be used to provide a cosmological context and choose initial conditions in numerical studies of black hole binaries assembled in star clusters. 
\end{abstract}



\section{Introduction} \label{sec:intro}
The detections of gravitational wave (GW) emission from ten distinct black hole binaries (BHBs) have conclusively demonstrated both that stellar BHBs form in nature, and that they can coalesce within a Hubble time \citep{ligo1, ligo2, ligo3, ligo4, ligo5,LIGO-Virgo2018}. However, dissipation of orbital energy and angular momentum via GW emission is only significant at very small separations ($\lesssim 10 R_{\odot}$ for circular orbits). While it is possible that binaries form with separations small enough for GW emission to result in coalescence within a Hubble time, such a formation scenario would require that the parent stars not expand significantly during stellar evolution, so as to avoid a stellar merger before both objects collapse into BHs \citep[e.g.,][]{mandel_mink_2016, mink_mandel_2016, marchant_etal_2016}. Alternatively, binaries may form with wide separations and shrink by a few orders of magnitude after a common envelope phase of stellar evolution \citep[e.g.,][]{dominik_etal_2012, ivanova_etal_2013, postnov_yungelson_2014,belczynski_etal_2016}. 

Yet another possibility, which does not depend on the details of stellar evolution, is that BHBs form in dense star clusters, such as in globular cluster (GC) progenitors \citep{morscher_etal_2015, rodriguez_etal_2015, rodriguez_etal_2016a, rodriguez_etal_2016b} or in the more massive nuclear star clusters \citep[NSCs;][]{portegies-zwart_mcmillan_2000, miller_lauburg_2009}. Because stellar densities are much higher in clusters than in the field, and because BHs segregate in the cluster core \citep{spitzer1969}, the rate of three-body interactions is enhanced significantly. As a result, BHBs can be assembled dynamically, either via direct three-body interactions \citep{lee1995} or via exchange interactions in close encounters with binary stars which may ultimately lead to the formation of a BHB \citep{goodman_hut_1993}. After their assembly, further hardening via binary-single scatterings provide a mechanism by which BHBs can contract to smaller separations. Depending on the magnitude of the velocity dispersion of the surrounding stars, the BHBs can either be ejected by dynamical recoil or be retained inside the cluster.

This dynamical formation channel has been explored extensively, both analytically \citep[e.g.,][]{lee1995,breen_heggie_2013a, breen_heggie_2013b} and numerically via Monte-Carlo \citep{gurkan2004, rodriguez_etal_2015, morscher_etal_2015, rodriguez_etal_2016a, rodriguez_etal_2016b, ar16, chatterjee_etal_2017a, chatterjee_etal_2017b, hong_etal_2018} and $N-$body simulations \citep{lee1995, oleary_etal_2006, miller_lauburg_2009, bae_etal_2014, park_etal_2017}. These studies have confirmed that GCs can indeed be the formation sites for a large population of BHBs (\citealt{rodriguez_etal_2016b}; see also \citealt{benacquista_downing_2013} for a review on the dynamical assembly of binaries).

Thus far, previous works have lacked detailed information about the initial conditions of the clusters being modeled. They typically adopt initial conditions based on observations of the Galactic GC population. However, GCs are expected to evolve significantly over $\sim$13 Gyr of cosmic history. Not only do the current properties of GCs differ from their state at formation, but also a large population of clusters that hosted BHB mergers may have been either disrupted by the local galactic tidal field \citep{gnedin_etal_1999, fall_zhang_2001, kravtsov_gnedin_2005, gieles_baumgardt_2008, gnedin_etal_2014} or spiraled into the central nuclear star cluster before the present day \citep[e.g.,][]{tremaine_etal_1975, capuzzo-dolcetta_miocchi_2008a, capuzzo-dolcetta_miocchi_2008b, antonini_etal_2012, antonini_2013, gnedin_etal_2014}. 

In this work, we provide constraints on the properties of the clusters most likely to be the nurseries of BHBs. To do so, we extract clusters from a cosmological model of GC formation and apply analytic estimates of relevant dynamical timescales. Our results are complementary to the study of BHB dynamics via more computationally intensive methods which are limited to modeling a small subset of the parameter space of initial conditions. 

\section{Methodology} \label{sec:methodology}

We begin this section with a brief outline of our GC formation and evolution model. Then we describe the cluster properties relevant to the evolution of BHBs and how we obtain each property (\autoref{subsec:methods1}). Given these cluster properties, we list our analytic prescriptions for the treatment of BHB formation and hardening (\autoref{subsec:methods2}). We finish with a description of the timescales over which the cluster may be disrupted (\autoref{subsec:methods3}). We adopt a flat, $\Lambda$CDM cosmology with $h=0.7$ and $\Omega_{\Lambda} = 0.7$. We use ``cosmic time'' to refer to the time from the big bang to a redshift $z$, and define the cosmic time at $z=0$ as $\thub \approx 13.7$ Gyr. Throughout, we refer to the initial mass of the cluster as $\Mcl$ and the mass at any other time as $\Mcl(t)$. 

\subsection{Determination of cluster properties}
\label{subsec:methods1}

We use the properties of GCs from the model of \citet[][hereafter, CGL18]{choksi_etal_2018}. The model assumes that GCs form in periods of rapid accretion onto dark matter halos (e.g., major mergers of galaxies). When such events occur, the total mass that forms in GCs scales linearly with the cold gas mass in the galaxy, which is in turn set by empirical galactic scaling relations. Clusters are drawn from a power-law cluster initial mass function $dN/d\Mcl \propto \Mcl^{-2}$ (see \citealt{choksi_gnedin_2018} for a discussion of the impact of the adopted mass function on the properties of GC systems) and are assigned metallicities based on the metallicity of their host galaxy using an empirical galaxy stellar mass-metallicity relation. GCs are evolved using prescriptions for both dynamical disruption and stellar evolution. The model was applied to $\approx$200 halo merger trees in the mass range $10^{11}\Msun \lesssim \Mh \lesssim 10^{15}\Msun$ from the \textit{Illustris} dark-matter-only simulation \citep{vogelsberger_etal_2014, nelson_etal_2015}. The resulting GC populations at $z=0$ are shown in CGL18 to match a wide range of the observed properties of GC systems. As a reference, we note that in the fiducial model, half of all clusters are predicted to form in the range $5<z<2.3$, corresponding to ages of $10.8-12.5$ Gyr, in halos of masses between $10^{11}-10^{12.5}\Msun$. These predictions are consistent with other recent GC formation models \citep{pfeffer_etal_2018, el-badry_etal_2018}.

We extract directly from the CGL18 model the initial cluster masses, $\feh$ values, formation times, and host galaxy properties for all clusters formed in the model. To relate $\feh$ to a total metallicity, we apply the conversion $\logz \approx \feh + 0.2$, based on the simulations of \cite{ma_etal_2016}. Below, we detail how we draw the remaining relevant cluster properties that the CGL18 model does not set.

The location of the cluster within the host galaxy is important because clusters may either inspiral due to dynamical friction and merge into the central nuclear star cluster or be gradually disrupted by the local tidal field before BHBs can be dynamically assembled. Observations of H$\alpha$ emission in high-redshift star forming galaxies show that most of the star formation occurs near the galaxy's half-mass radius, $R_{\rm h}$ \citep{forsterschreiber_etal_2018}. We therefore assume that GCs form with galactocentric radii distributed uniformly in the range (0.5-2)$R_{\mathrm{h}}$. 
We correlate $R_{\rm h}$ to the scale radius $R_{\mathrm{d}}$ of an exponential disk that has the same specific angular momentum as the dark matter halo. This results in a value of $R_{\mathrm{d}}\approx 2^{-1/2} \lambda\, R_{\mathrm{vir}}$, where $R_{\mathrm{vir}}$ is the virial radius and $\lambda$ is the dimensionless spin parameter of the host halo at formation \citep{mo_mao_white}. The scale length relates to the half-mass radius as $R_{\mathrm{d}} = 0.58\, R_{\mathrm{h}}$. We draw $\lambda$ from a log-normal distribution centered on $\lambda = 0.04$ with scatter $\sigma_{\lambda} = 0.25$ dex, typical for cosmological $\Lambda$CDM simulations \citep[e.g.,][]{rodriguez-puebla_etal_2016}. 

The initial size of the cluster will strongly affect the BHB merger rate \citep[e.g.,][]{morscher_etal_2015} because it sets the density, and in turn the BH interaction rate. Furthermore, GCs are expected to evolve in size due to relaxation effects and tidal stripping \citep[e.g.,][]{gnedin_etal_1999, gieles_baumgardt_2008, muratov_gnedin_2010}, so GC radii today cannot be used to estimate the initial properties of the cluster. Therefore, we turn to observations of ``young massive clusters'' in the local universe, which have masses and sizes consistent with objects that could evolve into GC-like systems after a few Gyr of dynamical and stellar evolution \citep{chatterjee_etal_2010, chatterjee_etal_2013, ymc_review, bastian2016}. These clusters show an approximately log-normal distribution in both their half-light radii $\rh$ (which we take as a proxy for the half-mass radius) and their core radii $\rc$, with peaks at $\rh \approx 2.8$ pc and $\rc \approx 1$ pc \citep{ymc_review, bastian_etal_2012, ryon_etal_2017}. We draw values of $\rc$ and $\rh$ from log-normal distributions centered on these values with standard deviations of 0.3 dex. Based on the observed correlation between $\rh$ and $\rc$ of young clusters, we also impose the requirement that $\rh \geqslant 2\rc$; if this condition is not met, we simply redraw $\rh$ until it is.

\begin{figure}
\includegraphics[width=\columnwidth]{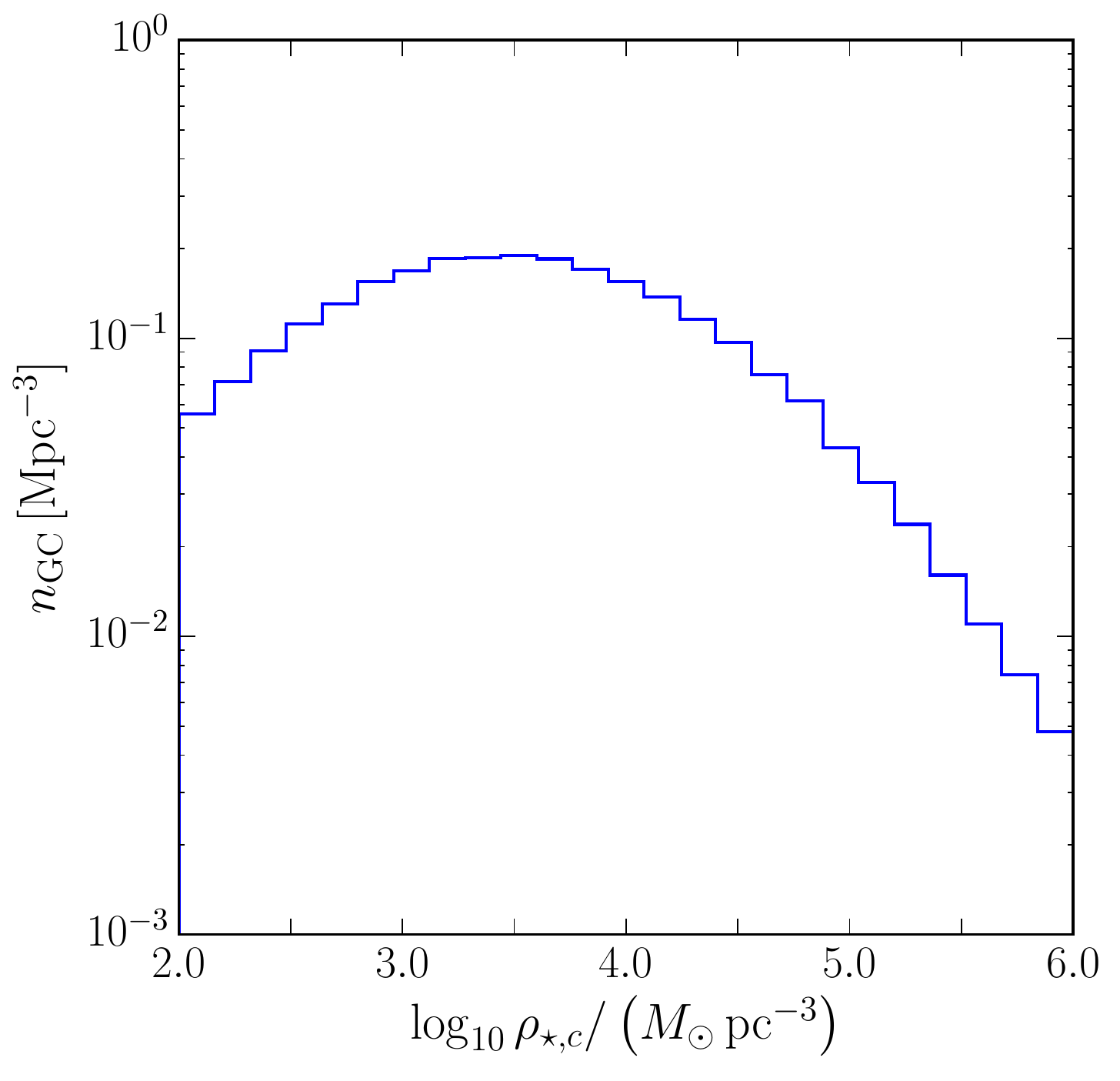}
\includegraphics[width=\columnwidth]{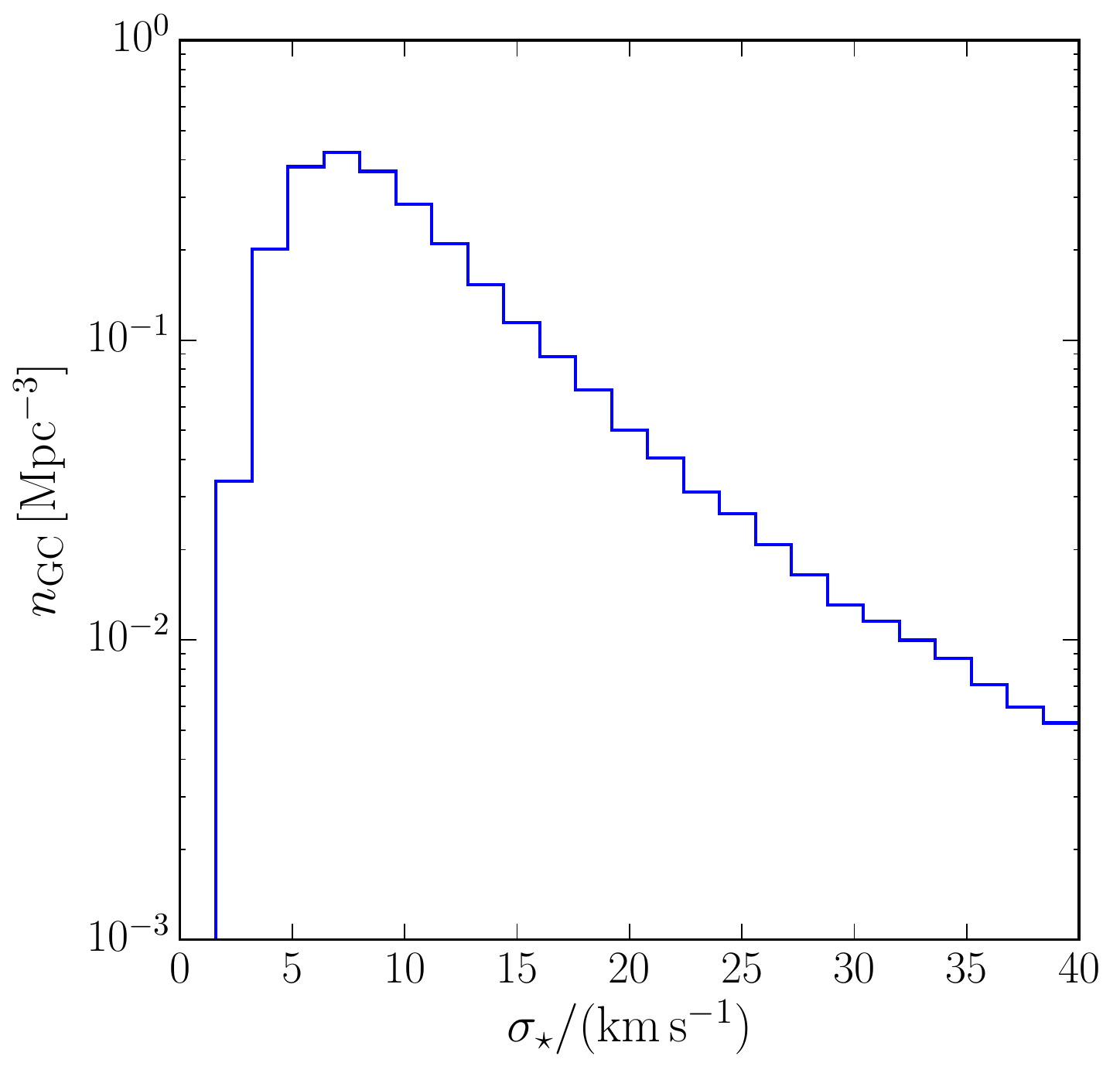}
\vspace{-5mm}
\caption{Distribution of central stellar densities, $\rhostarc$, and central velocity dispersions, $\sigstar$, calculated for all clusters extracted from the CGL18 model. The distributions are weighted by the halo mass function, so as to be cosmologically representative. The height of each bin gives the number density of clusters \textit{in the bin} (i.e., the total number density is given by the sum of bin values and not the integral under the histogram). }
  \label{fig:rho_sig_hists}
\end{figure}

We adopt the density and velocity profiles of \cite{stone_ostriker_2015} for all clusters. The profiles were designed to resemble single-mass King models, while also being analytically simple. They are fully described by three parameters: the half-mass radius $\rh$, the core radius $\rc$, and cluster mass $\Mcl$. The density of stars near the center of the cluster for this profile is then given by $\rhostarc = \Mcl(\rc + \rh)/(2\pi^2\rc^2\rh^2)$. Fig.~\ref{fig:rho_sig_hists} shows the distribution of stellar densities and velocity dispersions near the cluster centers for the adopted profiles and cluster sizes.

Two other relevant quantities are the average mass of a BH in the cluster, $\mbh$, and the fraction of the total cluster mass locked in BHs, $\Mbh/\Mcl$. We calculate both quantities for a \cite{kroupa_2001} stellar initial mass function (IMF), which has the form:
\begin{equation} 
\frac{dN}{dm_i}  \propto \begin{cases} m^{-1.3}_i &\text{if }  m_{\rm min} \leq m_i \leq 0.5 \Msun , \\ 
m_i^{-2.3} &\text{if } 0.5 \Msun < m_i < m_{\rm max}, \\
\end{cases}
\end{equation}
with lower and upper cutoff masses of $m_{\mathrm{min}} = 0.08 \Msun$ and $m_{\mathrm{max}} = 150 \Msun$. We have verified that our results are not sensitive to the particular choice of IMF. The IMF must be convolved with the relation between the mass of the progenitor star and compact remnant, $m_{\rm rem}(m_i, Z)$, which depends on metallicity due to line-driven winds in the final stages of stellar evolution \citep{woosley_etal_2002, belczynski_etal_2010a, belczynski_etal_2010b, spera_etal_2015, vink_2017}. For this, we adopt the results from the \textsc{sevn} code, which also accounts for the gap in the BH mass distribution in low metallicity environments due to pair instability supernovae \citep{spera_mapelli_2017}. We can then calculate $\Mbh$ and $\mbh$ as:
\begin{align}
\Mbh = \int_{m_{i, \rm min, \bullet}}^{m_{\rm max}} m_{\rm rem}(m_i, Z) \frac{dN}{dm_i} dm_i \nonumber \\
\mbh = \frac{\Mbh}{\int_{m_{i, \rm min, \bullet}}^{m_{\rm max}} \frac{dN}{dm_i} dm_i }.
\end{align}
The bounds of both integrals are taken from the minimum initial stellar mass that will ultimately form a BH  (set by the initial-final mass relation) to the upper mass cutoff of the IMF. In the following, for each cluster in the model $\mbh$ and $\Mbh$ are calculated based on its metallicity; as a reference, for our entire cluster population we find median values of $\mbh \approx 20 \Msun$ and $\Mbh/\Mcl \approx 5 \times 10^{-2}$.

\subsection{Formation and Coalescence of BHBs}
\label{subsec:methods2}

Below, we describe the time and spatial scales relevant to the formation of BHBs. Our prescription follows closely the approach of \citet[][hereafter AR16]{ar16}, who developed an analytic model for BHB evolution in star clusters. Unlike AR16, we do not perform a Monte Carlo draw of binaries in each cluster, but instead calculate typical quantities for the BHBs in each cluster, with the goal of understanding which types of clusters, on average, efficiently form merging BHBs. 

As the cluster evolves, the system seeks equipartition of kinetic energy, resulting in mass-segregation of the heavier BHs towards the cluster center on a timescale $\tau_{\rm MS} \approx \tau_{\mathrm{rlx}}{\mstar}/{\mbh}$, where $\tau_{\mathrm{rlx}}$ is the cluster relaxation time at the half-mass radius. Furthermore, if the cluster is susceptible to the Spitzer mass-stratification instability \citep{spitzer1969, watters_etal_2000}, the BHs  become confined to an ever-smaller core \rev{and form a decoupled self-gravitating sub-system. We find that all of our clusters are Spitzer-unstable.}

In reality, binary formation in the core provides an energy source that halts core collapse. BHB formation can occur through interactions between three BHs. The rate of BHB formation through this process is $\dot{n}_{\rm BHB} = 126G^5\mbh^5 n^3_{\bullet} \sigbh^{-9}$\rev{, where $n_{\bullet}$ and $\sigbh$ are the BH number density and velocity dispersion} \citep{goodman_hut_1993,lee1995}. The velocity dispersion and densities of BHs in the dynamically decoupled core can be related to the corresponding quantities for the background stars.
\rev{To do so, we begin by defining $\xi \gtrsim 1$ as the ratio of the mean kinetic energy of BHs to that of stars. This gives us the expression relating the stellar and BH velocity dispersions:
\begin{equation}
\sigbh = \left(\xi\frac{\mstar}{\mbh}\right)^{1/2}\sigstar.
\end{equation}
When the BHs dominate the mass in the core, the densities are related by \citep{lee1995}:
\begin{align}
\rhobhc &\approx \rho_{\rm h}\frac{\Mbh}{\Mcl}\frac{\rh^3}{r^3_{\bullet}} \nonumber \\ 
&\approx \frac{1}{2}\left(\frac{\rc}{\rh}\right)^2\left(\frac{\Mbh}{\Mcl}\right)^{-2}\rhostarc 
\label{eqn:core}
\end{align}
where $\rho_{\rm h}$ is the stellar density within $\rh$ and $r_{\bullet}$ is the half-mass radius of BHs. In computing the above expression, we assumed $\rh \gg \rc$ and that the BHs are confined within a radius $(\Mbh/\Mcl)$ smaller than the cluster half-mass radius. In reality, the density of BHs in the core evolves with time as mass segregation proceeds. Because our model includes no time evolution, we have here taken the limit that mass segregation is complete, which matches well results from $N-$body cluster simulations. }

Numerical simulations find that $\xi$ is typically a factor of a few \citep{gurkan2004}. We adopt $\xi = 5$, which gives timescales for BHB formation consistent with the numerical study of \cite{morscher_etal_2015}. Using these relations, we write the corresponding timescale for BHB formation as:
\begin{align}
\tthree &= 5\times 10^6 \, \left( \frac{\rho_{\star,\mathrm{c}}}{10^4 \Msun \mathrm{pc^{-3}}} \right)^{-2} \left(\frac{\mbh}{20\Msun} \right)^{-3} \left( \frac{r_{\mathrm{h}}/r_{\mathrm{c}}}{4} \right)^4  \nonumber \\
& \times \left(4\xi\frac{\mstar}{\mbh} \right)^{4.5} \left(\frac{\sigstar}{15\,\mathrm{km\,s^{-1}}} \right)^{9} \left( \frac{\Mbh/\Mcl}{10^{-2}} \right)^{4} \,\mathrm{yr}.
\label{eqn:tau_bhb}
 \end{align}

BHBs can also form by exchanging into stellar binaries that are born at the time of cluster formation or formed dynamically during cluster evolution. An estimate of the timescale for the formation of BHBs through single-binary exchanges is given by \citep{miller_lauburg_2009}: 
\begin{align}
\tsb = 3.75 \times 10^{9} \left(\frac{f_\mathrm{bin}}{0.1}\right)^{-1} \left( \frac{\rho_{\star, c}}{10^4\,\Msun\, \mathrm{pc^{-3}}} \right)^{-1} \left(\frac{\mstar}{1 \Msun}\right) \nonumber \\ \times \left(\frac{\sigstar} {\mathrm{15\,km\,s^{-1}}} \right) \left(4\xi\frac{\mstar}{\mbh}\right)^{1/2}\left(\frac{m_{123}}{20\Msun}\right)^{-1} \left(\frac{a_{\mathrm{hard}}}{1 \mathrm{AU}}\right)^{-1}\,\mathrm{yr}.
\label{eqn:t12}
\end{align}
Here $f_{\mathrm{bin}}$ is the stellar binary fraction, for which we adopt a value of 10\%, and $m_{123}$ is the mass of the stellar binary plus the BH, which we approximate as $2\mstar + \mbh$. Finally, $\ah$ is the typical separation of a hard stellar binary. We use as an estimate for $\ah$ the maximum separation of a hard stellar binary in the cluster core \citep{heggie_1975,quinlan_shapiro_1989}:
\begin{equation}
\ah = 1.5 \left(3\frac{q_{\star}}{1 + q_{\star}}\right)\left(\frac{r_{\mathrm{h}}}{3 \mathrm{pc}}\right)\left(\frac{\mstar/\Mcl}{10^{-5}}\right)\,\mathrm{AU},
\label{eqn:ahard}
\end{equation}
where $q_{\star} \leq 1$ is the mass ratio of the stellar binary. We adopt the average value of $q_{\star} \approx 0.5$ for a \cite{kroupa_2001} IMF. We define the binary formation timescale as $\tbhb = \mathrm{min(\tthree, \tsb)}$\footnote{We neglect the contribution of two-body binary formation through gravitational bremsstrahlung. \cite{lee1995} showed that this channel is only relevant for $\sigma \gtrsim \,100 \, \kms$, but typical GCs have $\sigma \sim 10 \, \kms$.}.

After BHBs form inside the cluster core, each BH-BHB interaction will cause them to both harden further and recoil. Denoting now the components of the BH binary $m_1$ and $m_2$, and the mass of the BH interloper $m_3$, the maximum separation of the binary below which interactions will lead to ejection from the cluster is \citep{ar16}:
\begin{align}
\aej = \frac{0.2G}{v^2_{\mathrm{esc}}} \frac{q_{12}}{(1 + q_{12})^2}\frac{q_3}{1+q_3}m_3 \nonumber \\
 = 0.26 \left(\frac{v_{\mathrm{esc}}}{30\, \kms} \right)^{-2} \left(\frac{m_3}{20 \Msun}\right) \nonumber \\ \times \left(5\frac{q_{12}}{(1 + q_{12})^2}\right)\left(3\frac{q_3}{1+q_3}\right)\,\mathrm{AU},
\label{eqn:aej}
\end{align}
where $q_{12} \leq 1$ is the mass ratio of the BHB, $q_3 \equiv m_3/(m_1 + m_2)$ and $v_{\mathrm{esc}}$ is the escape velocity from the cluster center. For a \cite{stone_ostriker_2015} potential-density pair, this is:
\begin{equation}
v_{\rm esc} = \frac{2}{\sqrt{\pi}}\sqrt{\frac{G\Mcl}{{\rh - \rc}}}\sqrt{\ln(\rh/\rc)}.
\end{equation}
\rev{The separation $\aej$ must be compared to the separation at which the BHB enters the regime in which gravitational waves dominate its evolution, $\agw$. A reasonable estimate for $\agw$ is obtained from setting the rate of dynamically induced shrinking of the binary \citep[][]{quinlan1996}:
\begin{equation}
\dot{a}_{\rm dyn} = -H\frac{G\rho}{\sigma^2}a^2,
\end{equation}
where $H \approx 20$ is the binary hardening rate, equal to the rate at which the binary shrinks due to gravitational wave emission \citep{peters1964}, $\dot{a}_{\rm GW}$. Using the core densities derived in Eq. \ref{eqn:core} yields the following expression for $\agw$:}
\begin{align}
\agw = 0.03 \left(\frac{m_1 + m_2}{40 \Msun}\right)^{3/5} \left(5\frac{q_{12}}{(1 + q_{12})^2}\right)^{1/5} \nonumber \\ \times \left(\frac{\sigstar}{15\, \kms}\right)^{1/5}\left(\frac{\rhostarc}{10^4 \Msun \mathrm{pc}^{-3}}\right)^{-1/5} \nonumber \\ \times \left(4\xi\frac{\mstar}{\mbh}\right)^{1/10}\left(\frac{\rh/\rc}{4}\right)^{2/5}\left(\frac{\Mbh/\Mcl}{10^{-2}}\right)^{2/5} f^{1/5}(e)\,\mathrm{AU}; \nonumber \\
f(e) = (1 + \frac{73}{24}e^2 + \frac{37}{96}e^4)(1 - e^2)^{-7/2},
\label{eqn:agw}
\end{align}
where $e$ is the eccentricity of the BHB, for which we adopt the average value $e = 2/3$ of a thermal distribution, $dP/de = 2e$ \citep{jeans_1919}. For each cluster we use the average values of $q_{12}$ and $q_3$, which we calculate based on the metallicity of each cluster, for a \cite{kroupa_2001} IMF convolved with the \cite{spera_mapelli_2017} initial-final mass relations. 

If $\agw < \aej$, the BHB will be ejected before it reaches the point that gravitational wave emission can quickly coalesce the BHBs. In this case, the BHB is ejected from the cluster with a separation $\aej$ and can continue to harden ex-situ only via GW emission. On the other hand, if $\agw > \aej$, GW emission will cause the BHBs to coalesce in-situ before dynamical ejection occurs.

Assuming that each interaction extracts 20\% of the binary's binding energy \citep[e.g.,][]{quinlan1996}, the time to harden to a separation $a_{\mathrm{crit}} = \mathrm{max}(a_{\mathrm{GW}}, a_{\mathrm{ej}})$, from an initial separation $a \gg a_{\rm crit}$, can be written as \citep{miller_hamilton_2002}\footnote{AR16 label this timescale $t_{\mathrm{merge}}$, but we find this label somewhat misleading and so opt for a different subscript.}:
\begin{align}
\tcrit = 2.5 \times 10^8 \left(\frac{q_3}{0.5}\right)^{-1}\left(\frac{\sigstar}{15\,\kms}\right)\left(4\xi\frac{\mstar}{\mbh}\right)^{1/2} \nonumber \\ \times \left(\frac{\acrit}{0.1 \rm AU}\right)^{-1}\left(\frac{m_1 + m_2}{40 \Msun} \right)^{-1}\left(\frac{\rh/\rc}{4}\right)^2 \nonumber \\ \times   \left(\frac{\Mbh/\Mcl}{10^{-2}}\right)\left(\frac{\mbh}{20 \Msun}\right)\left( \frac{\rhostarc}{10^4 \Msun \mathrm{pc}^{-3}}\right)^{-1} \, \mathrm{yr}.
\label{eqn:tacrit}
\end{align}

After reaching $\acrit$, the binary's hardening is dominated by the emission of GWs, which drives the system to coalescence on a timescale \citep{peters1964}:
\begin{align}
\tgw = 2.8 \times 10^{8} \left(\frac{m_1m_2(m_1 + m_2)}{10^4 \Msun^3}\right)^{-1} \left(\frac{\acrit}{0.1 \mathrm{AU}}\right)^4 \nonumber \\ \times  \left(\frac{1-e^2}{0.5} \right)^{7/2}\,\mathrm{yr}.
\label{eqn:tgw}
\end{align}

\subsection{Timescales for cluster disruption}
\label{subsec:methods3}

In this subsection, we introduce the timescales over which a cluster may be destroyed. In Section \ref{sec:results}, these will be compared to the timescales for the formation and evolution of BHBs.

\begin{figure}
\includegraphics[width=\columnwidth]{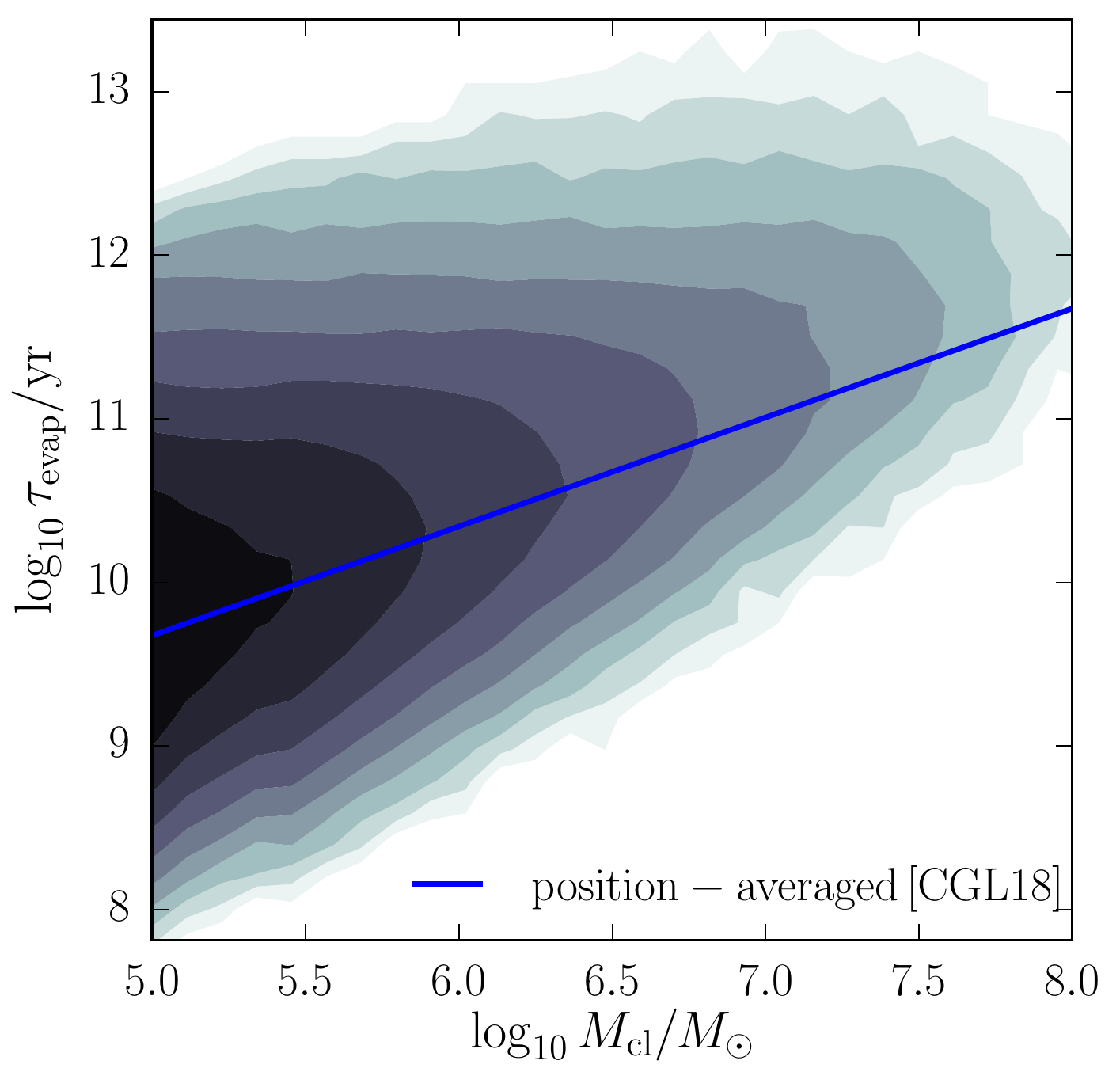}
\vspace{-5mm}
\caption{Contours show the distribution of cluster disruption times as a function of the cluster mass $\Mcl$, taking into account their galactocentric radii. The blue line shows the position-averaged disruption time used in CGL18, which reproduces the $z=0$ cluster mass function.}
  \label{fig:tdis}
\end{figure}

Two-body relaxation in star clusters leads to gradual ``evaporation" of stars from the cluster and can eventually lead to the complete dynamical disruption of the cluster. To model this effect, we adopt the prescription of \cite{gnedin_etal_2014} for mass loss in the presence of a strong external tidal field\footnote{For simplicity, we ignore evaporation in the weak tidal field limit, which is almost always subdominant.}:
\begin{equation}
\frac{dM}{dt} = -\frac{\Mcl(t)}{\ttid(\Mcl, R)},
\label{eqn:disruption1}
\end{equation}
where $R$ is the galactocentric radius of the GC and $\ttid$ is given by \citep{gieles_baumgardt_2008}:
\begin{align}
\ttid(\Mcl,R) = 10^{10} \left(\frac{\Mcl(t)}{2 \times 10^5 \Msun}\right)^{2/3}P(R) \,\mathrm{yr}\\
P(R) \equiv 41.4 \left(\frac{R}{\mathrm{kpc}}\right)\left(\frac{V_{\mathrm{circ}}}{\mathrm{km\,s^{-1}}}\right)^{-1}.
\label{eqn:disruption2}
\end{align}
Here $V_{\mathrm{circ}}$ is the circular velocity in the galactic potential. We assume a flat rotation curve, with $V_{\mathrm{circ}} = \sqrt{0.5GM_{\mathrm{bar}}/R_{\mathrm{h}}}$,  and $M_{\mathrm{bar}} = \Mstar + M_{\mathrm{gas}}$ is the total mass in baryons of the galaxy where the GC formed, which is also extracted from the CGL18 model. Integrating Eq.~\ref{eqn:disruption1} from the initial mass of the cluster, $\Mcl$, to $M=0$, gives the time to complete evaporation of the cluster:
\begin{equation}
\tdis = 1.5 \times 10^{10}\left(\frac{\Mcl}{2 \times 10^5 \Msun}\right)^{2/3}P(R)\,\mathrm{yr}.
\label{eqn:tdis}
\end{equation}
CGL18 ignored the variation in spatial position of clusters, and instead adopted an average value of the factor $P(R)$ which allowed them to match the $z=0$ GC mass functions. We find that our prescription for assigning cluster positions reproduces the average disruption timescale adopted by CGL18 well (Fig.~\ref{fig:tdis}). 

Finally, the cluster may inspiral into the center of the galaxy due to dynamical friction. We do not integrate the orbit of each cluster within the host galaxy's potential, but instead use an approximate dynamical friction timescale as calculated by \citet{binney_tremaine}, evaluated at the initial position and mass of the cluster:
\begin{align}
\tdf \approx 2.3 \times 10^{10} \left(\frac{\Mcl}{2 \times 10^5 \Msun}\right)^{-1} \left(\frac{R}{\mathrm{kpc}} \right)^2 \nonumber \\ \times \left(\frac{V_{\mathrm{circ}}}{100\,\mathrm{km\,s^{-1}}}\right)\, \mathrm{yr}.
\label{eqn:tdf}
\end{align}
After a time $\tdf$ elapses we assume the cluster is tidally disrupted by, and subsequently merges with, the NSC. Finally, we define the timescale for the disruption of the cluster, $\tdes = \mathrm{min(\tdis, \tdf)}$.

The above prescriptions for the evolution of the GC within the host galaxy potential are necessarily simplifications. For example, GCs may migrate out of the disk and into the stellar halo after a galaxy merger \citep[e.g.,][]{kruijssen2015, li_etal_sim3} and gradual mass loss of the cluster during its inspiral may increase the total inspiral time \citep{gnedin_etal_2014}. Nevertheless, we believe these prescriptions should provide a reasonable approximation on average.

\section{Results} \label{sec:results}

\begin{figure}
\includegraphics[width=\columnwidth]{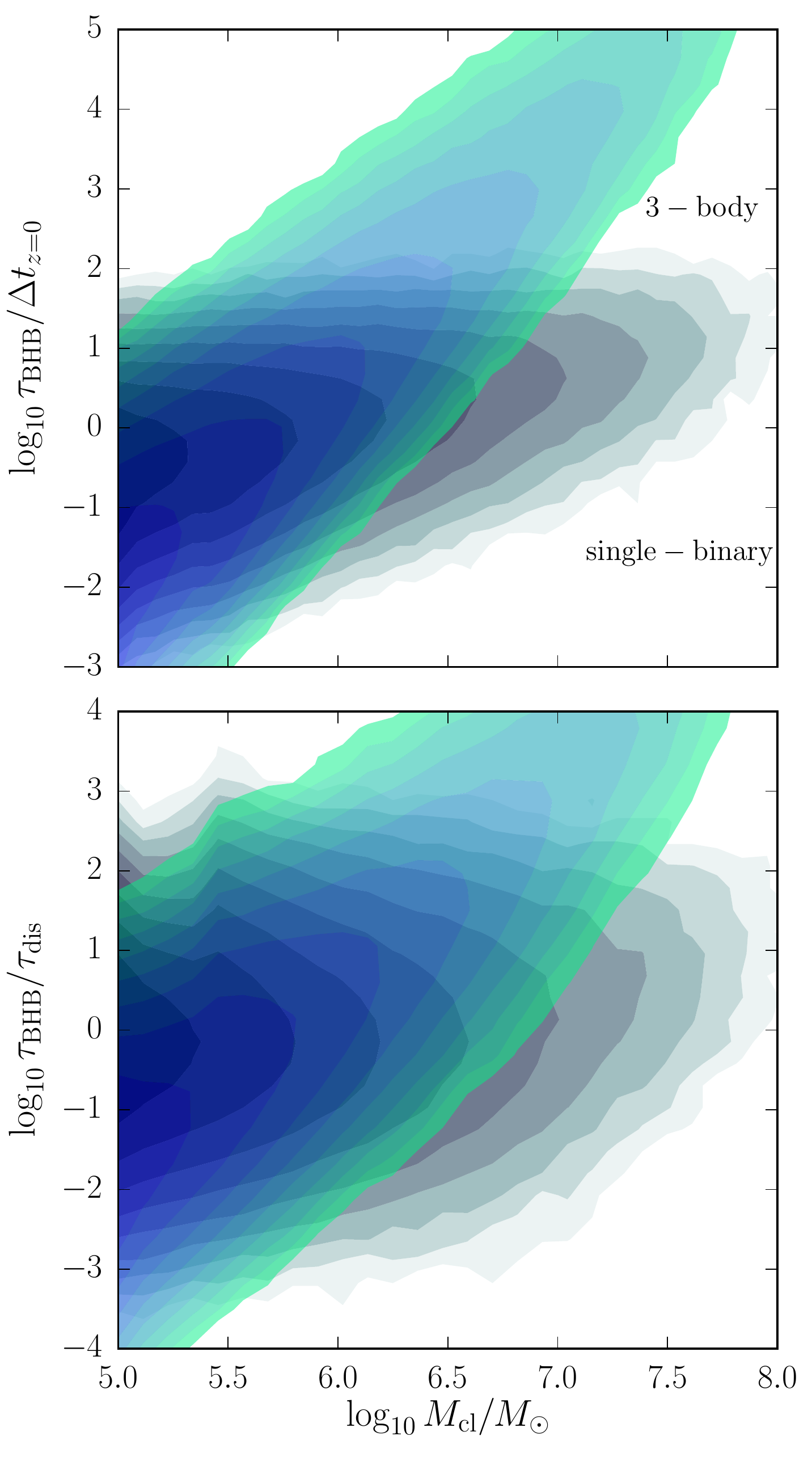}
\vspace{-5mm}
\caption{Timescale for formation of BHBs via 3-body and single-binary exchange, normalized by $\Delta t_{z=0} \equiv t(z=0) - t(z_{\mathrm{form}})$ (upper panel) and the cluster disruption time $\tdes \equiv \mathrm{min(\tdis, \tdf)}$ (lower panel), as a function of the cluster mass, $\Mcl$.}
  \label{fig:tbhb}
\end{figure}

\begin{figure}
\includegraphics[width=\columnwidth]{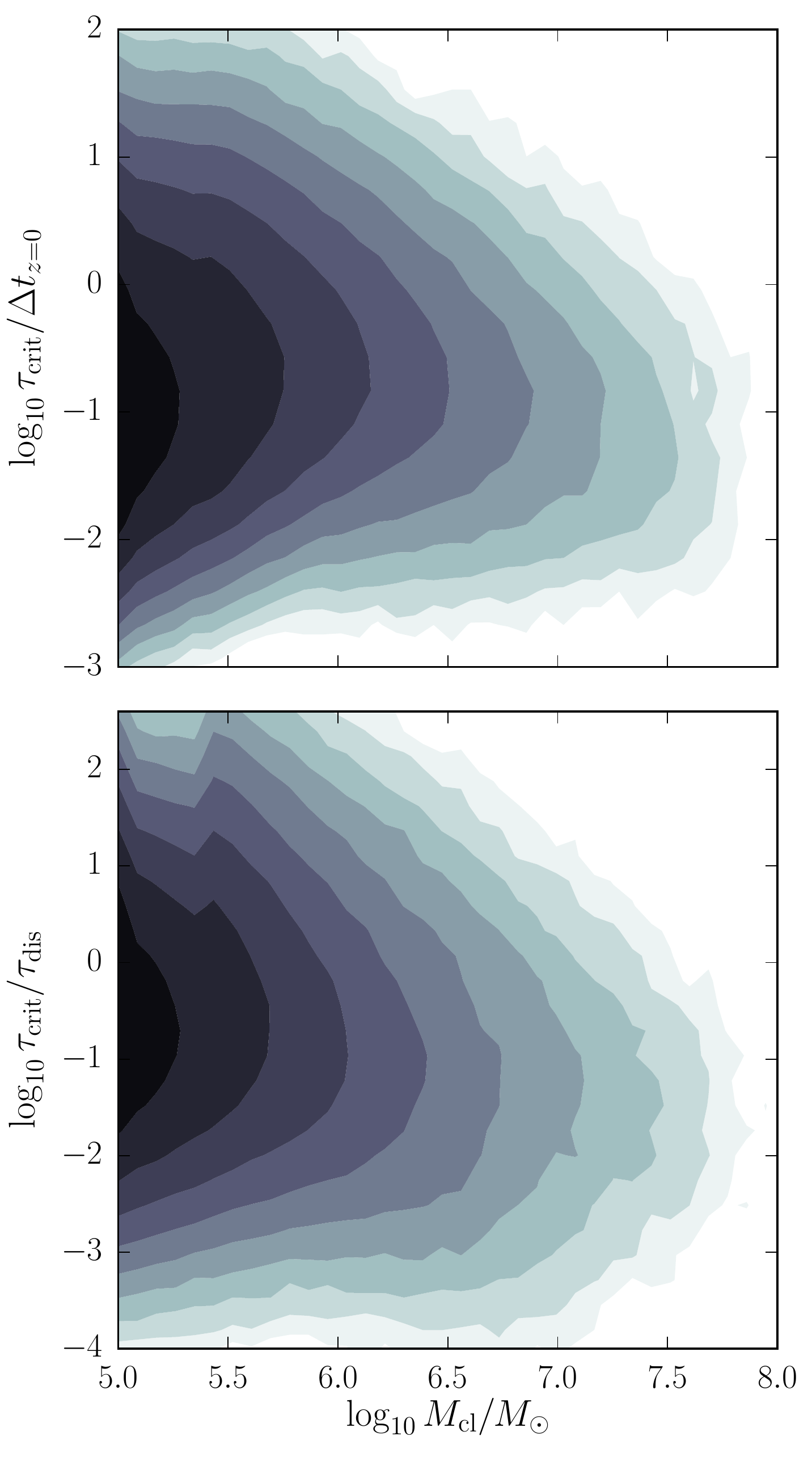}
\vspace{-5mm}
\caption{Same as Fig.~\ref{fig:tbhb}, but now showing the timescale to harden the binary separation to $\acrit = \mathrm{max(\agw, \aej)}$.}
  \label{fig:tcrit}
\end{figure}

\begin{figure}
\includegraphics[width=\columnwidth]{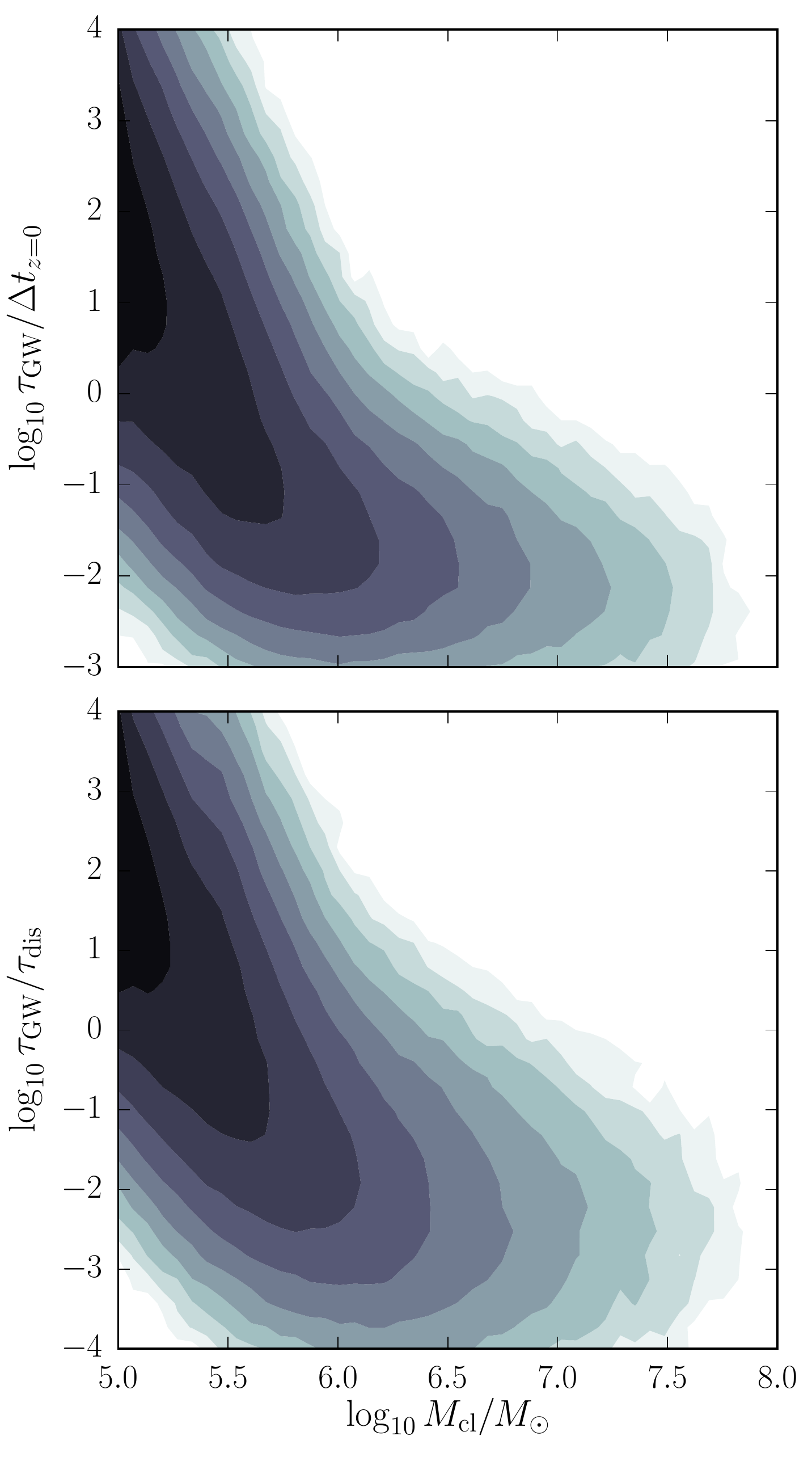}
\vspace{-5mm}
\caption{Same as Figs.~\ref{fig:tbhb}-\ref{fig:tcrit}, but now for the timescale to merge via GW emission, evaluated at $\acrit$. The change in the scaling of $\tgw$ with $\Mcl$ at $\Mcl \sim 10^6 \Msun$ occurs because $\agw$ begins to exceed $\acrit$ at this mass (see Fig.~\ref{fig:agw_aej_percentiles}).}
\label{fig:tgw}
\end{figure}

\begin{figure}
\includegraphics[width=\columnwidth]{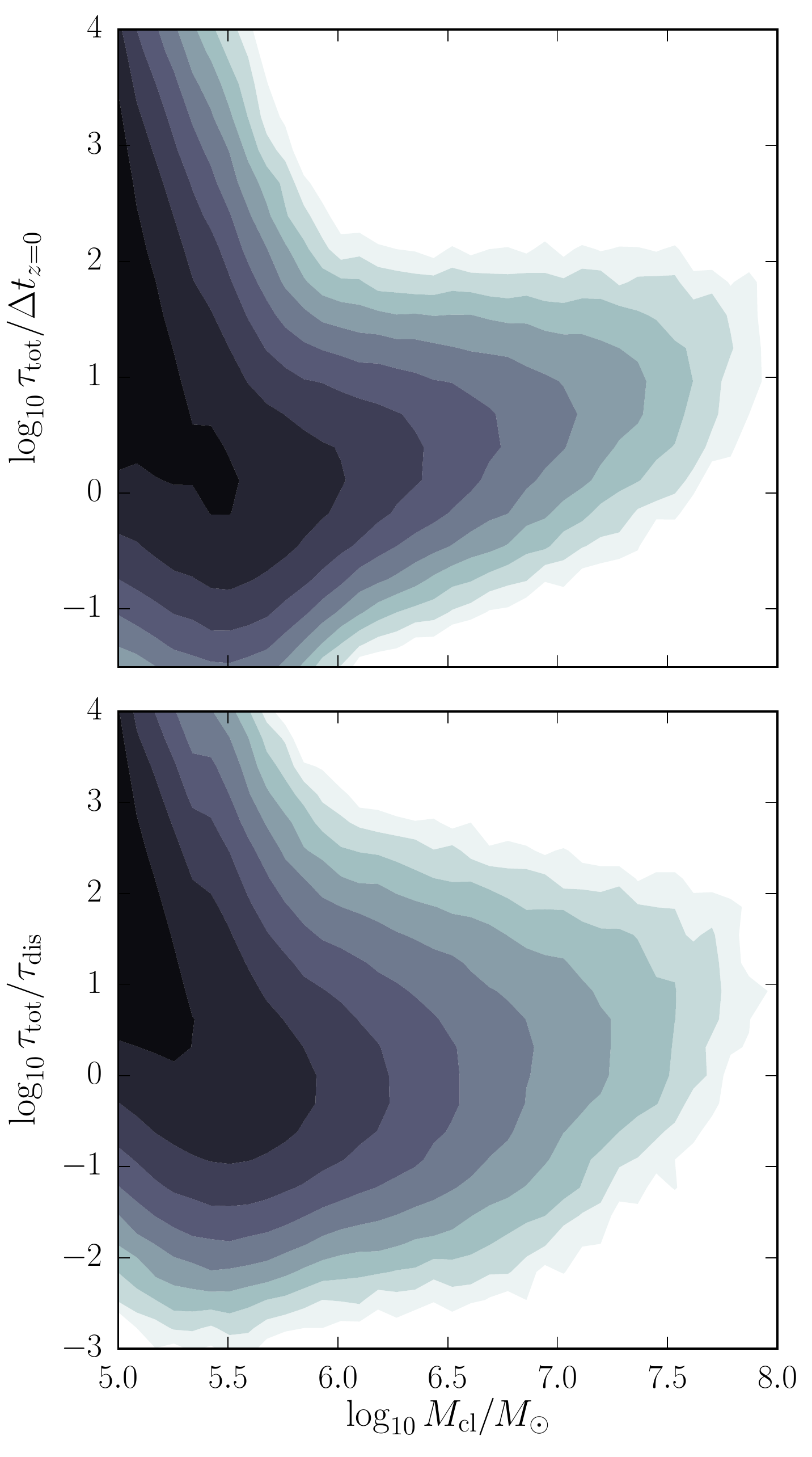}
\vspace{-5mm}
\caption{Total time to BHB mergers for all clusters (see Eq.~\ref{eqn:ttot}).}
  \label{fig:ttot}
\end{figure}

\begin{figure}
\includegraphics[width=\columnwidth]{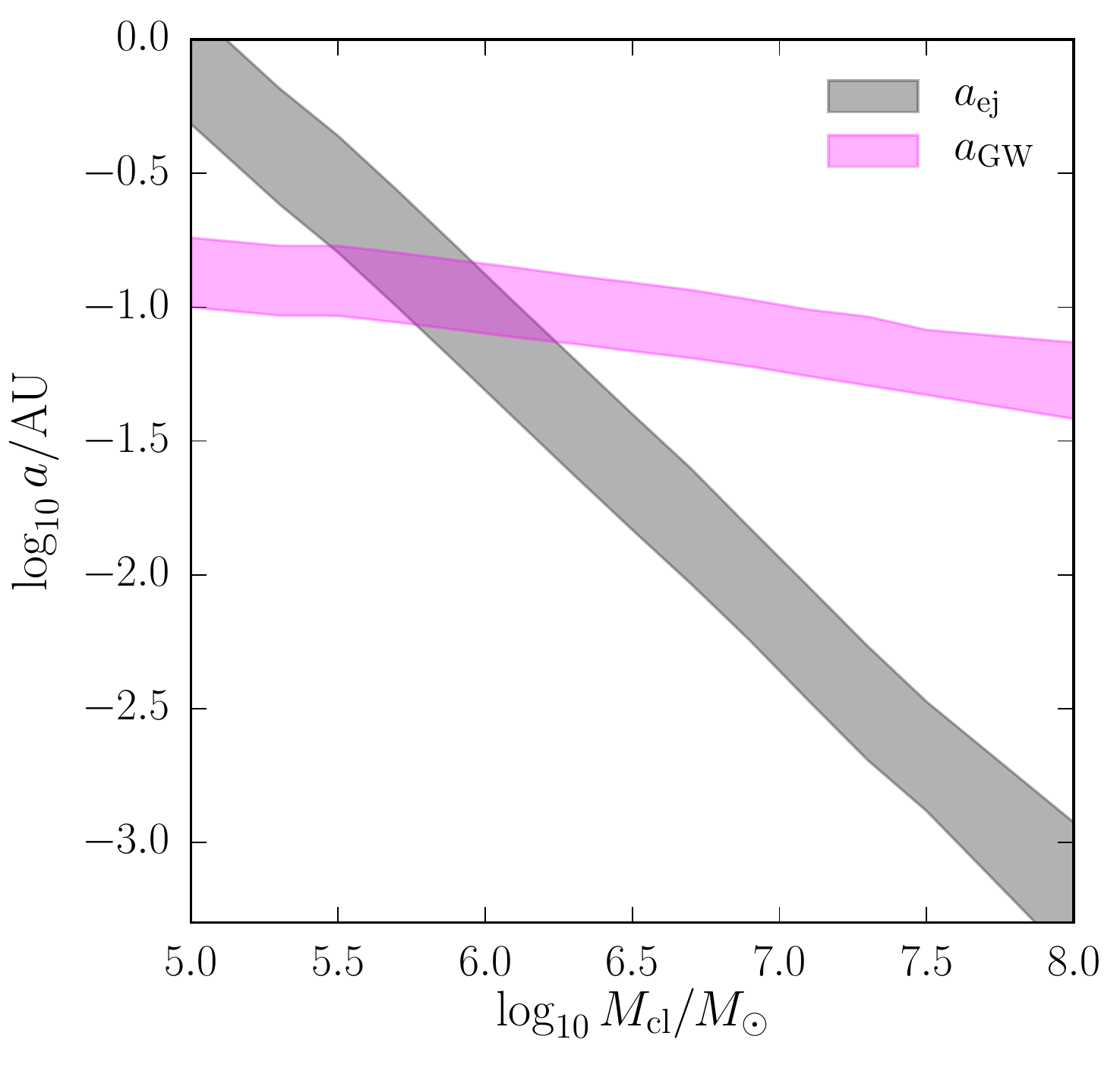}
\caption{$\aej$ and $\agw$ as a function of $\Mcl$. The shaded regions show interquartile ranges (25th to 75th percentile). To the left of the intersection, $\aej > \agw$ and BHBs merge ex-situ, while to the right, $\aej < \agw$ and BHBs merge in-situ. The transition mass between the two regimes is $\Mcl \approx 10^{6} \Msun$.}
\label{fig:agw_aej_percentiles}
\end{figure}

\begin{figure}
\includegraphics[width=\columnwidth]{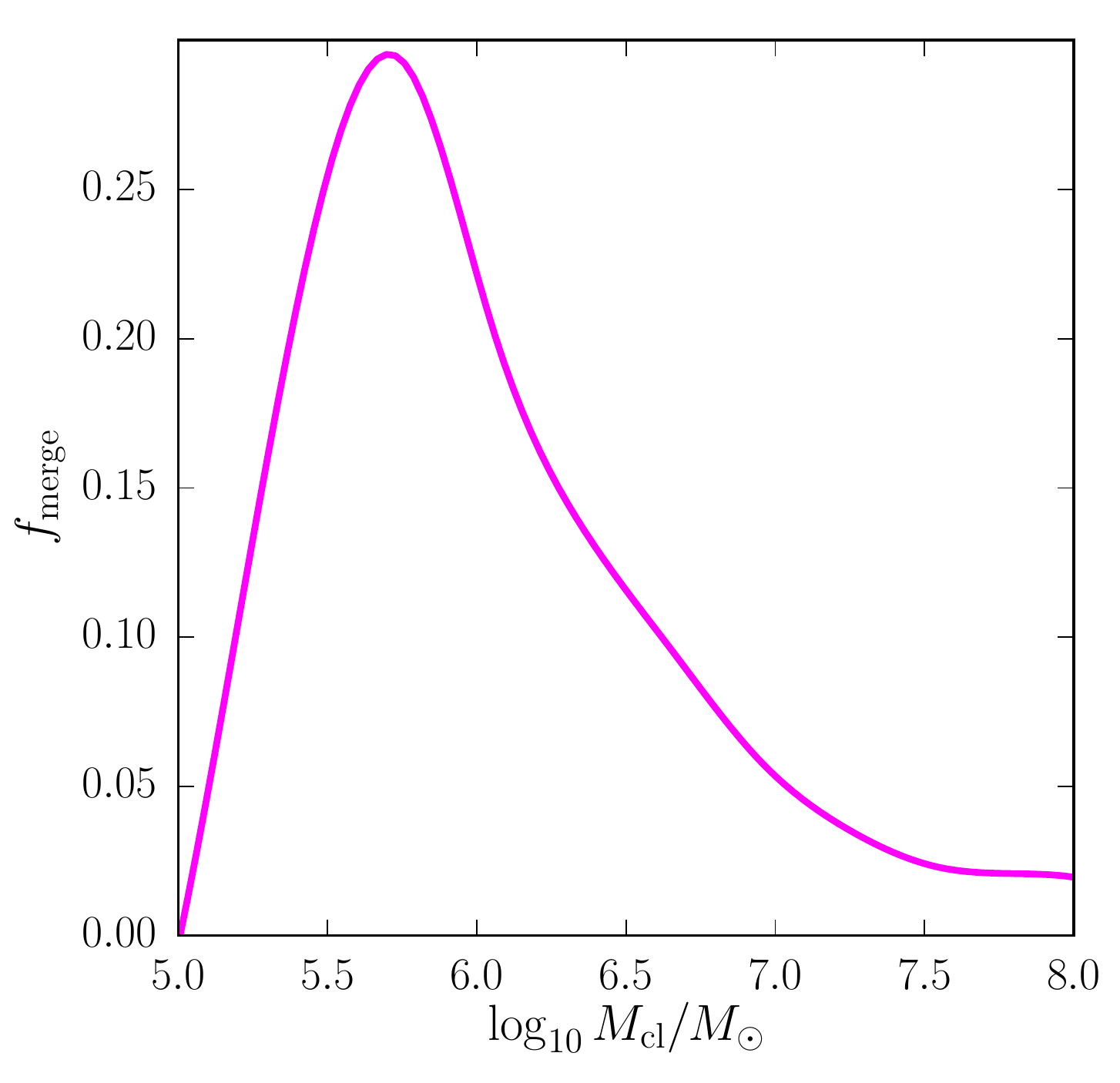}
\caption{The fraction of clusters, at fixed cluster mass, in which BHBs merge efficiently. \rev{This is calculated by dividing our cluster sample into narrow bins of cluster mass and calculating the fraction of clusters in each bin that satisfy the criteria put forth in equations \ref{eqn:conditions} and \ref{eqn:conditionstwo}.} The peak occurs near $\Mcl \approx 10^{5.7} \Msun$, around which 30\% of clusters in that mass range produce merging BHBs.}
  \label{fig:fmerge}
\end{figure}

\begin{figure*}
\includegraphics[width=\textwidth]{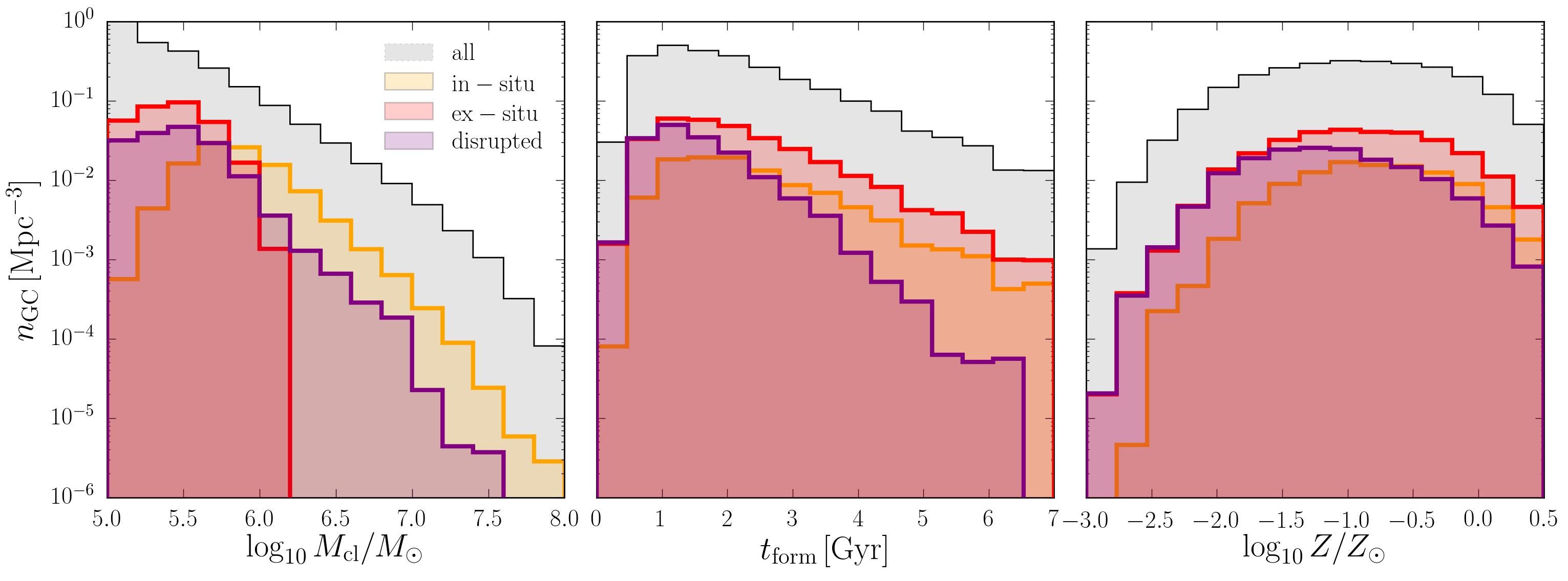}
\vspace{-5mm}
  \caption{Distributions of cluster masses, cosmic time at formation, and metallicities. Yellow and red show clusters where BHB mergers occur in-situ and ex-situ, respective. Purple shows clusters that both disrupt before $z=0$ and whose BHBs merge (either in or ex-situ). The distributions are weighted by the halo mass function, so as to be cosmologically representative. The height of each bin gives the number density of clusters in the bin (i.e., the total number density is given by the sum of bin values and not the integral under the histogram). The total number density of clusters for ``all'', ``in-situ'', ``ex-situ'', and ``disrupted'' are 2.6 $\Mpcc$, 0.10 $\Mpcc$, 0.31 $\Mpcc$, and 0.16 $\Mpcc$. }
  \label{fig:merging_hists}
\end{figure*}

In this section, we begin by presenting the distributions of the various timescales discussed in \autoref{sec:methodology}. Then, we use a combination of these timescales to specify criteria for identifying the clusters most likely to produce merging BHBs. We then analyze the properties of these clusters relative to the overall cluster population. We conclude with an estimate of the cosmic BHB merger rate.

\subsection{Timescales}

The typical time for a BHB in a given cluster to merge is:
\begin{equation}
\ttot = \tau_{\mathrm{MS}} + \tbhb + \tau_{\mathrm{crit}} + \tgw.
\label{eqn:ttot}
\end{equation}
Figs.~\ref{fig:tbhb}-\ref{fig:ttot} show the various timescales\footnote{The mass-segregation timescale, $\tau_{\rm MS}$, is always short compared to $\thub$ ($\lesssim 10^8$~yr), so we do not show it here.} as a function of $\Mcl$, normalized by both $\Delta t_{z=0} \equiv t(z=0) - t(z_{\mathrm{form}})$ (i.e., the cosmic time between cluster formation and $z=0$) and the disruption timescale, $\tdes$.

\rev{For clusters with mass $\Mcl \lesssim 10^{5.5} \Msun$, 
the binary formation timescale is set by the three-body interaction timescale because of the low velocity dispersions in these systems (Fig. \ref{fig:tbhb}). At higher masses ($\Mcl \sim 10^{5.5} - 10^{6} \Msun$), three-body interactions and single-binary exchange both contribute, while at the very highest masses the timescale for single-binary exchange to pair BHs is always much shorter than the three-body timescale. Although single-binary exchange dominates at high cluster masses, the timescale for it to form BHBs is comparable to $\dt$. Thus, binary formation is difficult at high cluster masses.}

Fig.~\ref{fig:tcrit} shows the timescale $\tcrit$ over which BHBs dynamically shrink in separation to $\acrit$. The median time spent in this phase is roughly constant at $\sim$10\% of $\dt$. After a time $\tcrit$ has elapsed, BHBs spend a time $\tgw$ in the GW dominated regime, which is shown in Fig.~\ref{fig:tgw}. For the lowest mass clusters with $\Mcl \sim 10^5 \Msun$, we find that $\tgw$ is generally very large compared to $\dt$ because BHBs are ejected from the cluster with large separations, since these clusters have low escape speeds. More massive clusters, on the other hand, can retain and dynamically harden BHBs until they reach the regime where hardening via GW emission is effective, and thus $\tgw$ is very short for typical BHBs in these clusters. The vastly different scalings of $\tgw$ with $\Mcl$ are caused by the switch from evaluating $\tgw \propto \acrit^4$ at $\aej \propto \Mcl^{-1}$ to $\agw \propto \Mcl^{-1/10}$, which occurs at a typical mass of $\Mcl \sim 10^{6} \Msun$ (Fig.~\ref{fig:agw_aej_percentiles}).

The scaling of $\ttot$ with cluster mass changes as different components of $\ttot$ begin to dominate (Fig. \ref{fig:ttot}). At low masses, $\ttot$ is typically greater than $\dt$ and dominated by $\tgw$. At intermediate and high masses ($\Mcl \gtrsim 10^{5.5} \Msun$), $\ttot$ is dominated by BHB formation.

\begin{table*}
\begin{center}
\begingroup
\hspace{-10ex}\begin{tabular}{lll}
\hline
Description & Percent of all clusters\\
 \hline
BHB mergers (in or ex-situ) & 15.8\% \\ 
BHB mergers ex-situ & 11.8\% \\ 
BHB mergers in-situ & 4.0\% \\ 
BHB mergers (in or ex-situ) + disrupt before $z=0$ & 6.3\% \\
\hline
\end{tabular}
\endgroup
\\
\end{center}
\caption{Summary of the fraction of clusters that satisfy various conditions. The middle two rows split all clusters that form merging BHBs into those in which the BHBs merge ex-situ ($\aej > \agw$) and in-situ ($\agw > \aej$). The final row refers to clusters that disrupt, either via dynamical evaporation or a merger with the NSC, but also produce BHBs that merge (regardless of where the BHB mergers occur).}  
\label{tab:fractions}
\end{table*}

The time $\ttot$ does not solely determine whether a cluster can form BHBs that will merge before $z=0$. If $\agw > \aej$, the BHBs will merge in-situ, and it is therefore necessary that the cluster not disrupt before BHB mergers can occur. Thus, we require:
\begin{align}
\ttot < \Delta t_{z=0} \nonumber  \\
\ttot < \tdis \nonumber \\ 
\ttot < \tdf.
\label{eqn:conditions}
\end{align}
On the other hand, if $\agw < \aej$, BHBs will merge ex-situ. In this case, the destruction of the cluster will not affect the mergers of its BHBs, because they have already been ejected from the cluster. Thus, the conditions for a cluster to produce BHBs that merge before $z=0$ are slightly relaxed compared to Eq.~\ref{eqn:conditions}:
\begin{align}
\ttot < \Delta t_{z=0} \nonumber  \\
\ttot - \tgw < \tdis \nonumber \\ 
\ttot - \tgw < \tdf,
\label{eqn:conditionstwo}
\end{align}
where $\ttot - \tgw$ represents the time to form and harden BHBs to a separation $\acrit$.

In both of the above sets of conditions, we implicitly ignore the contribution of clusters which disrupt before their BHBs have shrunk in separation to $\acrit$, even though hardening via GW emission will continue and may coalesce BHBs after the cluster disrupts and its constituents join the field or an NSC. For less massive clusters, where $\acrit = \aej$, BHBs with separation $a \gtrsim \aej$ would have GW coalescence times orders of magnitude greater than $\dt$ because $\tgw \propto a^4$. For more massive clusters, where $\acrit = \agw$, in reality some BHBs may shrink to a separation $a \gtrsim \agw$ which is small enough that GW emission would be sufficient to merge the BHBs after cluster disruption. Thus, we may slightly underestimate the number of clusters that merge BHBs in-situ. However, based on the fraction of clusters that disrupt, our estimate can be off by no more than a factor of two.

\subsection{Properties of clusters that make merging BHBs}

\rev{We find that $\approx$15\% of all clusters produce BHBs that can merge before $z=0$. This fraction differs from some recent estimates derived by extrapolating results from numerical simulations, which find a much higher fraction of clusters producing merging BHBs. The difference is driven by the fact that our model spans a much wider range in initial cluster sizes than can be sampled in numerical simulations. In Section \ref{subsec:merger_rate}, we discuss in detail the implication of larger cluster sizes on the cosmic BHB merger rate.}

In Fig.~\ref{fig:fmerge} we divide the model clusters into bins of cluster mass and show the fraction of clusters in each bin that form merging BHBs. The curve peaks at a value of $\approx$30\% for $\Mcl \approx 10^{5.7} \Msun$, with a decreasing fraction at both lower and higher masses. Thus, clusters near a mass of $10^{5.7} \Msun$ are the most effective, relative to the overall population at the same mass, at forming merging BHBs. 

The median mass of all the clusters that form merging BHBs is $\Mcl = 10^{5.5} \Msun$, which is 0.25~dex higher than the median value for the entire cluster population. We can further divide into clusters that eject BHBs before coalescence and those that do not. The first panel in Fig.~\ref{fig:merging_hists} shows the mass distribution for the two cases, as well as the distribution for all clusters for comparison. In the case of ex-situ BHB mergers, the number of clusters peaks and is about constant in the range $10^5 \Msun -10^{5.5} \Msun$, with a sharp truncation at higher masses. For the in-situ case, the number of clusters peaks at $\Mcl \approx 10^{5.7} \Msun$. Approximately three times more clusters will have ex-situ BHB mergers than in-situ. This is largely due to the fact that in-situ mergers happen in clusters that are a few times rarer, because of the bottom-heavy cluster initial mass function, $dN/d\Mcl \propto \Mcl^{-2}$. We note that if BHBs are on circular orbits, in contrast to our adopted eccentricity value $e = 2/3$, then the ratio of ex-situ to in-situ mergers will increase.

These distributions can be understood as follows. As previously discussed, for low-mass clusters ($\Mcl \sim 10^5 \Msun$), $\aej \gg \agw$ and BHBs are ejected from the cluster with large separations and correspondingly long $\tgw$. As a result, only a small fraction of low-mass clusters form BHBs that can merge, leading to the nearly flat distribution of cluster masses for the case of ex-situ mergers, even though low-mass clusters are the most numerous. At intermediate masses ($\Mcl \sim 10^{5.3} \Msun$), $\aej \gtrsim \agw$, and although clusters typically eject BHBs before they reach $\agw$, the ejection separation is small enough that the BHBs can coalesce before $z=0$ solely via gravitational wave emission after ejection. At higher masses ($\Mcl \gtrsim 10^{5.7} \Msun$), $\agw \gtrsim \aej$, and the cluster escape velocities are high enough that a large fraction of clusters can merge BHBs in-situ. Finally, at the highest cluster masses, BHB formation timescales are comparable to $\dt$ (see Fig.~\ref{fig:tbhb}), so BHB formation and mergers are rare, thus leading to the slow decrease in the number of successful clusters out to the highest cluster masses. 

The distribution of the cosmic times of formation for clusters that form merging BHBs peaks at $t \approx 1.5$ Gyr, with a long tail extending towards later formation times. Unlike the distribution of formation times, the metallicity distribution is roughly flat between $-2 < \logz < 0$. The wide spread in metallicity reflects the fact that globular cluster systems assemble from a diverse range of host galaxies which later merge to combine their respective GC populations. 

The metallicities and formation times of clusters that form merging BHBs are similar to that of the overall cluster population. The ratio of the median metallicity of clusters whose BHBs merge to that of the overall population, $\log_{10}Z_{\rm merge}/Z_{\rm all}$, decreases with increasing $\Mcl$ and varies between +0.3~dex at $\Mcl \approx 10^5 \Msun$ and -0.2~dex at $\Mcl \approx 10^8 \Msun$. Likewise, at $\Mcl \approx 10^5 \Msun$ clusters whose BHBs merge typically form 0.4 Gyr later than the overall population at that mass. The offset decreases with increasing $\Mcl$ and vanishes for $\Mcl \gtrsim 10^6 \Msun$. The net result across all cluster masses is that the median metallicity for clusters that form merging BHBs is essentially identical to that of the overall population, and the median formation time is 0.1 Gyr later.

Clusters that evaporate or merge with the NSC before $z=0$ also contribute to the BHB merger rate, despite the fact that disruption limits the available time for BHB formation and hardening (see Fig.~\ref{fig:merging_hists}). Approximately 40\% of the clusters that produce merging BHBs will also disrupt before $z=0$. The majority of these disrupt due to dynamical evaporation, while only a small fraction merge with the NSC. For comparison, 60\% of all clusters formed in the model are disrupted before $z=0$. 

The distributions of properties of the clusters who both disrupt and form BHBs that eventually merge are represented by the purple curves in Fig.~\ref{fig:merging_hists}. The similar heights of the curves for the disrupted clusters (purple) and the clusters whose BHBs merge ex-situ (red) in the mass range $10^5 \Msun - 10^{5.5} \Msun$ demonstrates that a large fraction of the hosts of ex-situ merging BHBs will have been disrupted by $z=0$. In contrast, almost all of the hosts of in-situ merging BHBs will still be around at the present-day. 

The distributions of core and half-mass radii of clusters whose BHBs merge also follow log-normal distributions, but these distributions are shifted to smaller radii by $\approx$0.3 dex relative to the adopted distributions for all clusters. The distribution of formation loci in the galaxy is indistinguishable from that of the whole population. 

Table~\ref{tab:fractions} summarizes the fraction of clusters satisfying various conditions.

\subsection{Cosmic BHB merger rate} \label{subsec:merger_rate}

While the focus of this paper is to derive the properties of the clusters where BHBs are likely to form, using the framework outlined above for determining whether a cluster produces BHBs that merge by redshift $z$, we can provide a rough estimate of the cosmic merger rate of dynamically assembled BHBs as a function of redshift or cosmic time. 

To estimate the merger rate, we select all clusters in the CGL18 model that satisfy the conditions in equations \ref{eqn:conditions}-\ref{eqn:conditionstwo}. For each successful cluster we assume that one binary forms and coalesces after a time $\ttot$ has elapsed from the time of the cluster formation. Having the merger trees and masses of the halos where the clusters form gives us the number density of GCs as a function of cosmic time. We estimate the weight of each halo using its number density, $W_{\rm HMF}$, from the halo mass function of the parent cosmological $N-$body simulation. With the information on the time when binaries merge we can directly calculate the rest-frame intrinsic merger rate by summing the number of BHB mergers per cosmic time: 
\begin{equation}
\frac{dn_{\mathrm{com}}}{dt} = \sum^{N_{\rm {tree}}}_{k=1} \frac{W_{\rm HMF}}{N^{(k)}_{\rm h}} \frac{N^{(k)}_{\mathrm{mergers}}}{dt},
\label{dndtdV}
\end{equation}
where $N^{(k)}_{\mathrm{mergers}}$ is the number of mergers within a time $dt$, $N^{(k)}_{\rm h}$ is the number of $z=0$ halos in each bin, and $N_{\rm tree}$ is the total number of merger trees available, or equivalently, the total number of $z=0$ halos \citep[similar to the approach taken to estimate the merger rate for more massive black holes in the {\it LISA} band in][]{arun_etal_2009}. 

The number of events per unit observation time over the full sky is then given by:
\begin{equation}
\frac{dN}{dt}=\int
4\pi c \,\frac{dn_{\rm com}(z)}{dz}\left[\frac{D_L(z)}{(1+z)}\right]^2\, dz,
\label{dNdt}
\end{equation}
where $n_{\rm com}(z)$ is the comoving density of events at a given redshift (analogue to Eq.~\ref{dndtdV} but in redshift rather than time bins) and $D_L$ is the luminosity distance \citep{haehnelt1994}.

To account for multiple BHBs forming in a given cluster, we note that most clusters which can form merging BHBs have $\Mcl \sim 10^{5.5} \Msun$ (see Fig. \ref{fig:merging_hists}). These clusters typically have $\feh \approx -0.7$, which corresponds to $\Mbh/\Mcl \approx 3 \times 10^{-2}$ and $\mbh \approx 20 \Msun$. Thus, a cluster of this mass will have $\approx$ 240 BHBs, assuming all BHs are paired. While not all BHs become merging BHBs, we ignore in this calculation any cluster for which the average BHB does not merge before $z=0$ (following the various timescale conditions). In reality, these clusters will also contribute to the merger rate. We therefore multiply our lower limit by 240 to obtain a more realistic estimate of the merger rate, and the results can be rescaled accordingly. However, we caution that this calculation represents only a rough estimate of the normalization of the merger rate; other trends such as the dependence on halo mass or redshift are more robust. We also do not apply a signal-to-noise threshold since we only use the average mass of BHs in each cluster. 

Including this correction factor, we obtain at $z=0$ an intrinsic merger rate of 5.8 $\gpcyr$, similar to those estimated in previous work. From run O1 of LIGO the rate of BHB mergers is in the range between 12 and 213 $\gpcyr$ within the $90\%$ credible region. Thus  $>10\%$ of LIGO-Virgo events might have a dynamical origin. The total event rate, per unit of observation time, from Eq.~\ref{dNdt} is $\approx 1975 \, \mathrm{yr}^{-1}$.

\begin{figure}
\vspace{2mm}
\includegraphics[width=\columnwidth]{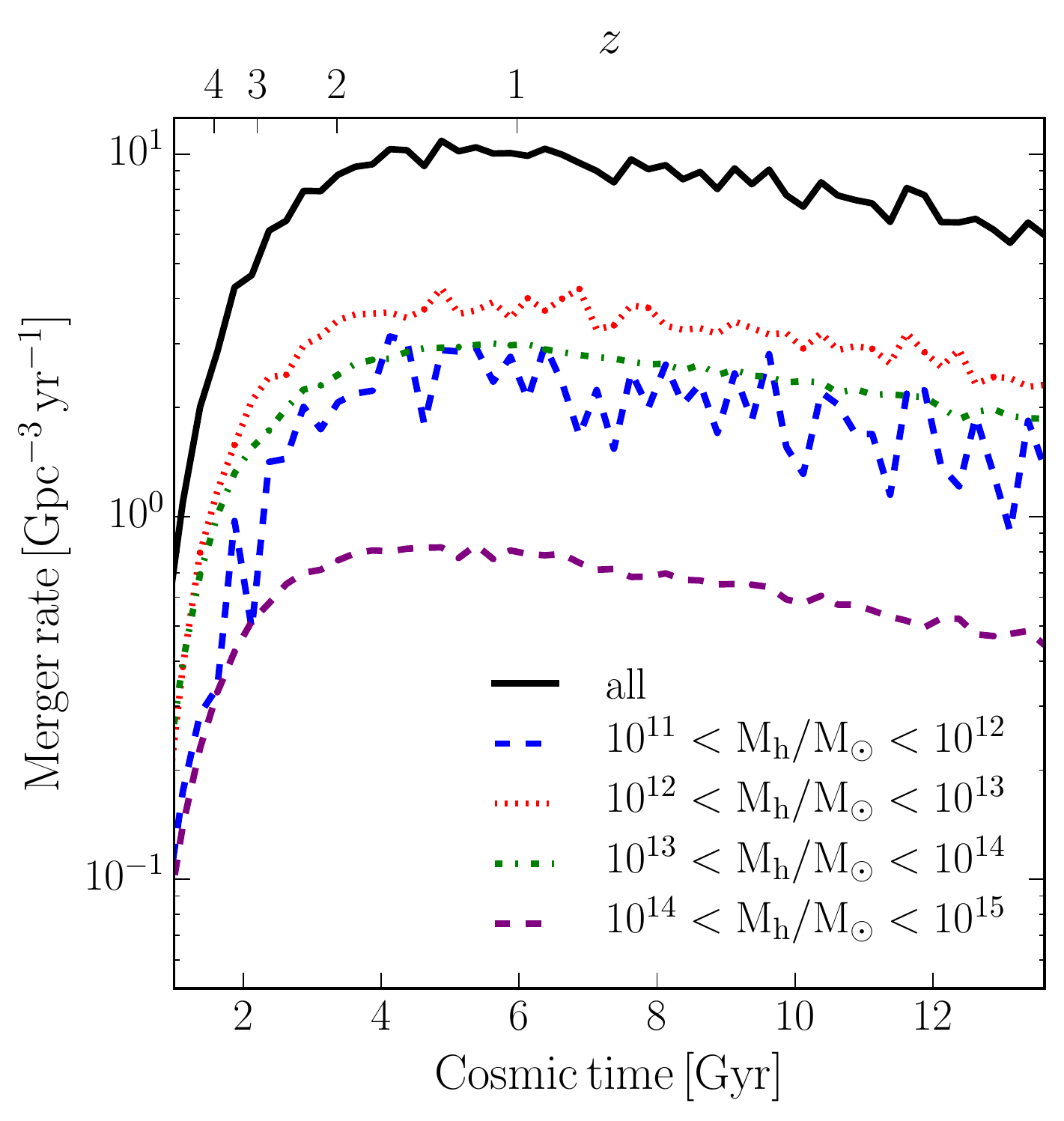}
\vspace{-4mm}
\caption{Predicted intrinsic BHB merger rate per comoving volume as a function of redshift. The black solid curve shows the total rate, while the colored, dashed curves split the merger rate in bins of halo mass. We emphasize that the normalization of the merger rate here is only a rough approximation, assuming 240 BHBs merge per cluster satisfying the conditions in \ref{eqn:conditions}--\ref{eqn:conditionstwo}, and that all other clusters do not contribute BHB mergers. The factor 240 comes from assuming that all clusters producing merging BHBs are of mass $\Mcl = 10^{5.5} \Msun$ (see Fig.~\ref{fig:merging_hists}).}
  \label{fig:ligo_z}
\end{figure}

Fig.~\ref{fig:ligo_z} shows our predicted intrinsic BHB merger rate as a function of cosmic time. After splitting into bins of halo mass and weighting by their respective number densities, we find that halos in the range $10^{12} \Msun \lesssim \Mh(z=0) \lesssim 10^{13} \Msun$, similar to or slightly more massive than the Milky Way, have the largest contribution to the merger rate when averaging over a large, unbiased cosmic volume. The most massive halos, with $\Mh \gtrsim 10^{14} \Msun$, always contribute the least. Their low contribution is due to their rarity, not because their clusters have significantly different properties. Conversely, low-mass halos, $\Mh \lesssim 10^{12} \Msun$, contribute less than Milky Way-sized halos because they have fewer globular clusters, although the halos themselves are more frequent. However, because the number of successful clusters \textit{per halo} scales linearly with the $z=0$ halo mass, in the case of a specific gravitational wave event which is localized to a patch of sky, the most massive halos would be the most likely hosts of the merger, if the binary was dynamically assembled in a star cluster.

\subsection{Comparison to other merger rate estimates}
\begin{figure}[t]
\vspace{-2mm}
\includegraphics[width=\columnwidth]{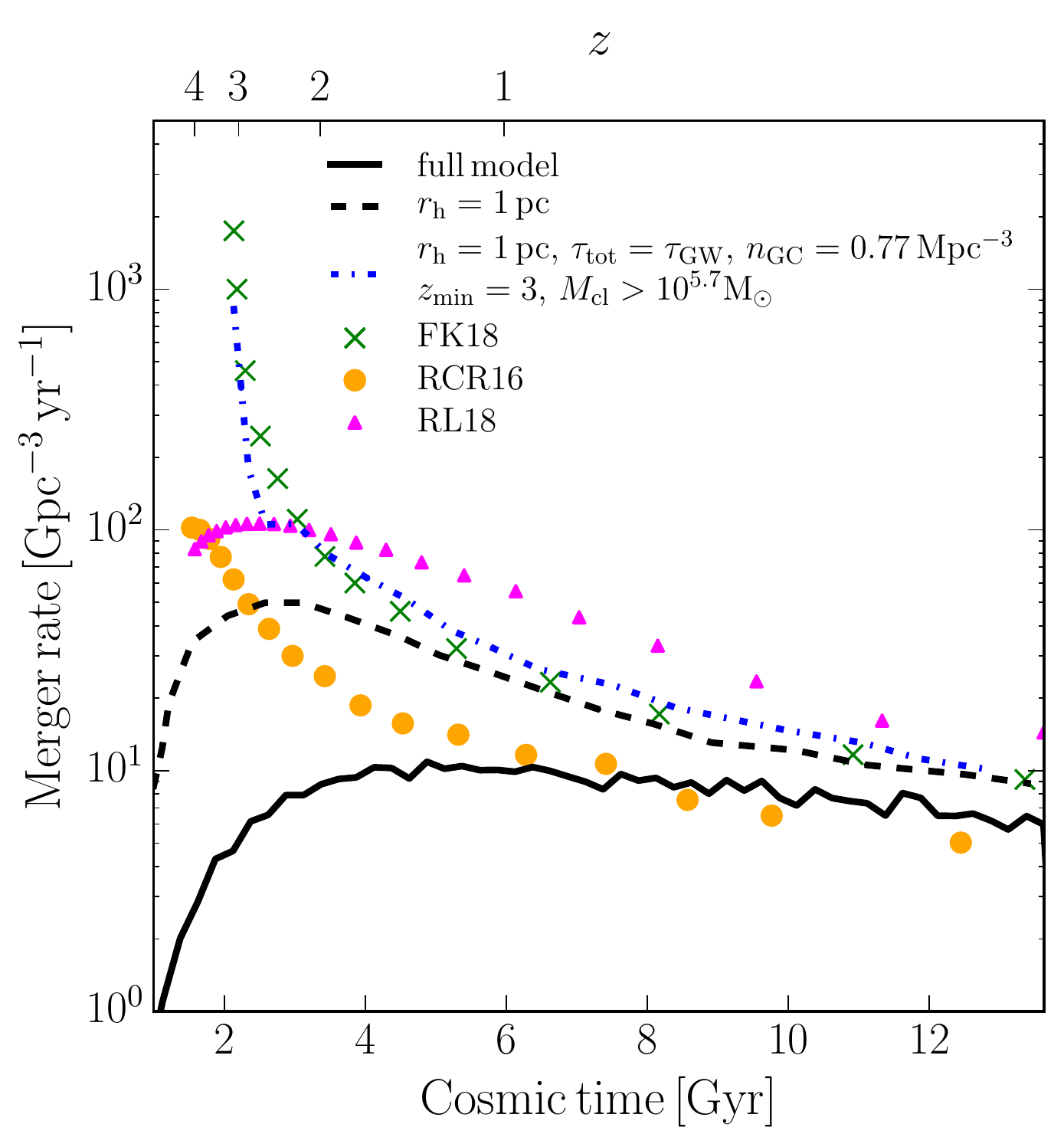}
\vspace{-6mm}
\caption{Comparison of the merger rate with previous estimates. We report as a black solid curve the same curve of Fig.~\ref{fig:ligo_z}, i.e., our full result in which cluster sizes are set from log-normal distributions peaked at $\rh = 2.8$ pc and $\rc = 1$ pc. The dashed line shows the model result when we fix $\rh = 1$ pc and $\rc = 0.5$ pc. Overplotted as points are the rates from the $r_{\rm vir} = 1$ pc model of \cite{rodriguez_loeb_2018} and the fiducial model of \citet{rodriguez_etal_2016b}. \rev{The blue shaded region bounds the lower and upper limits of the rates calculated by \cite{fragione_kocsis_2018}.}}
  \label{fig:ligo_z_comp}
\end{figure}

The redshift evolution of the estimated merger rate can be compared to those predicted by other works \citep{rodriguez_etal_2016b, fragione_kocsis_2018, rodriguez_loeb_2018}. We find a weakly increasing rate from $z=0$ out to $z=1.5$ ($t(z=1.5)=4.3$ Gyr), and a steep drop at higher redshift. \cite{rodriguez_etal_2016b} calculated the merger rate evolution by assuming all clusters formed at $z=4$ and using the $z=0$ number density and mass function of GCs combined with the distribution of merger delay times from their suite of Monte Carlo cluster simulations. \rev{\cite{fragione_kocsis_2018} also used fits to the same set of numerical simulations and assumed all clusters formed at $z=3$, taking into account also the evolving number of GCs due to disruption using the GC evolution model of \cite{gnedin_etal_2014} of a single Milky Way-type halo.} To compute the total BHB merger rate, they extrapolated their results to all galaxies. Both works found a steady increase in the merger rate with increasing redshift out to their starting epoch. Finally, \cite{rodriguez_loeb_2018} used GC formation rates from the cosmological GC formation model of \cite{el-badry_etal_2018} combined with delay times from the \cite{rodriguez_etal_2016b} simulations. This study is most similar to ours, and finds a qualitatively similar redshift evolution but with a peak at higher redshift, $z \approx 3$. \rev{In Fig. \ref{fig:ligo_z_comp} we compare the various predicted merger rates.}

A major difference between all three of these models and ours is the assumed cluster sizes. The suite of simulations ran by \cite{rodriguez_etal_2016b}, upon which all three models rely in calculating the BHB merger delay times, only cover clusters with sizes of 1 and 2 pc. Consequently, they find that BHB formation and hardening in the cluster occur quickly, and that the bottleneck is the time spent in the GW-dominated regime. 

\begin{figure}
\includegraphics[width=\columnwidth]{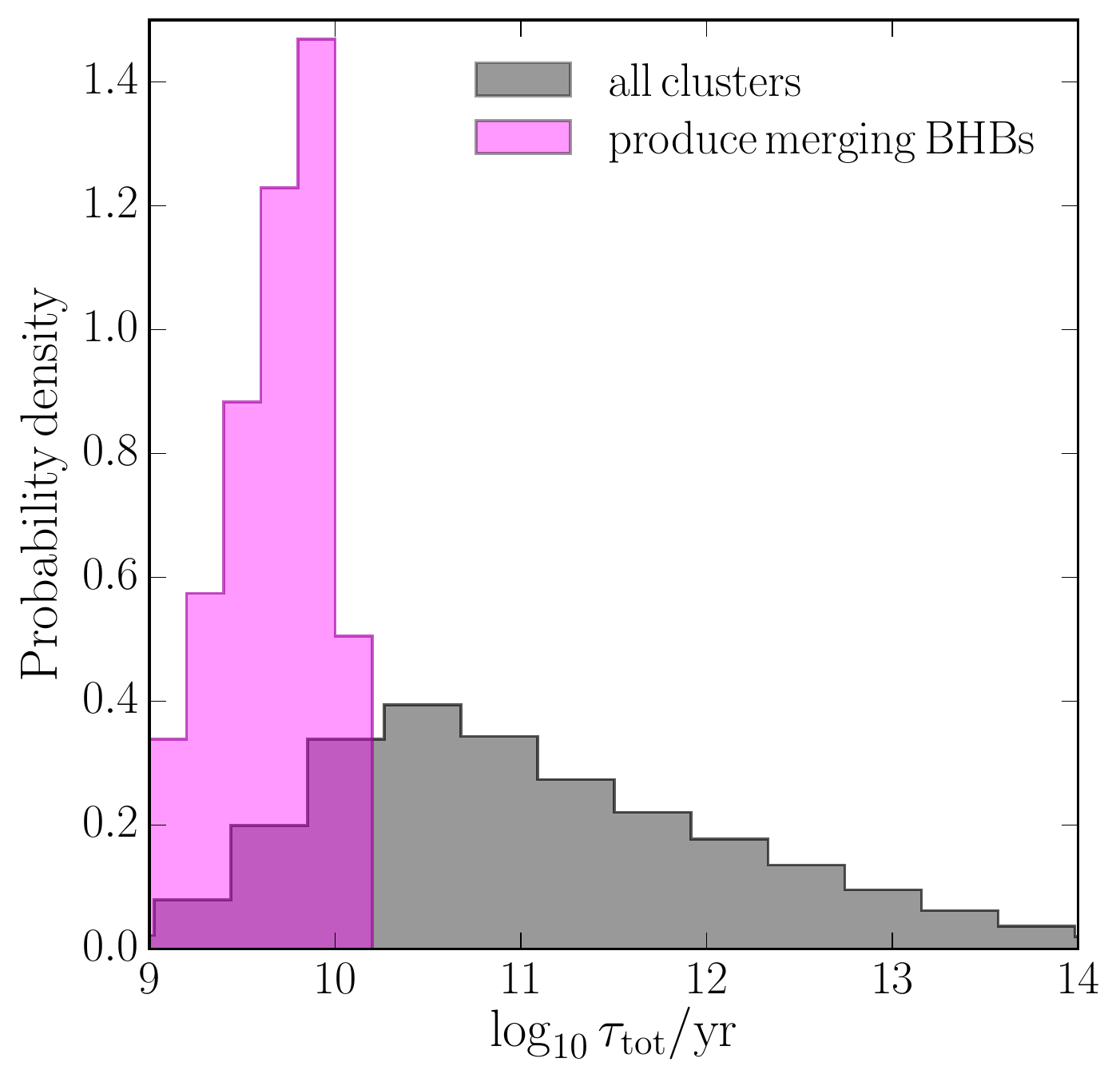}
\caption{Distributions of the delay times to BHB mergers, from the time of cluster formation (i.e., $\ttot$ in our model). The distribution is skewed towards long delay times, resulting in the prediction of Fig. \ref{fig:ligo_z} that the cosmic BHB merger rate of dynamically assembled BHBs peaks at relatively low redshift, $z \approx 1.5$. The integral under each histogram is normalized to unity.}
  \label{fig:ttot_hist}
\end{figure}

\rev{In contrast, because our model accounts for the full range of cluster sizes, we find that dynamical processes introduce in most clusters a non-negligible delay before BHBs reach the GW regime. Consequently, the typical delay time between cluster formation and BHB mergers in our fiducial model is a few Gyr, pushing the peak of the merger rate to lower redshift and lowering the overall normalization (Fig. \ref{fig:ttot_hist}). Using fixed values of $\rh = 1$ pc and $\rc = 0.5$ pc for our entire cluster population, we find a nearly identical redshift evolution to that predicted by RL18. There remains a factor of $\approx$2 difference in the normalization of the merger rate between our $\rh = 1$ pc model and the RL18 model at all redshifts; the discrepancy is likely due to a combination of effects relating to differences in the adopted GC formation model and treatment of BHB dynamics as well as the simplifying assumptions we made in calculating the merger rate. Given that the technique used to estimate the merger rate is approximate (since this is not the focus of this paper) we defer a more complete discussion of possible differences to future work aimed at this purpose (Choksi et al., in prep).} 

\rev{The redshift evolution of our model is also similar to those of \cite{rodriguez_etal_2016b} and \cite{fragione_kocsis_2018} at $z \lesssim 3$, but differs at higher redshift. Both these models assume all GCs form in a single burst at their starting epoch; using an extended GC formation history removes the initial peak in the merger rate predicted by their models and causes a turnover, as seen in Fig. \ref{fig:ligo_z}. Finally, we note that at $z=0$ all recent predictions of the merger rate agree to within a factor of a few, irrespective of the assumed cluster disruption rate or initial conditions. At higher redshift, the various predictions diverge by up to an order of magnitude. }

\section{Conclusions} \label{sec:conclusions}

In this work we analyzed the properties of the star clusters most likely to produce merging BHBs. Our model is based on realistic cluster populations extracted from a cosmological model of GC formation \citep{choksi_etal_2018}. To each of these clusters, we applied analytic prescriptions for the formation, hardening, and merging timescales of BHBs, following the method outlined by \cite{ar16}. This method allowed us to determine which clusters were likely to form merging BHBs and whether or not BHB mergers were likely to happen in the cluster. We extend on previous work by modeling a diverse and realistic population of GCs, rather than a small subset of the parameter space. Our main conclusions are:
\begin{enumerate}
\item We confirm previous calculations that massive star clusters can effectively create and shrink in separation black hole binaries. $\sim$15\% of all clusters can produce BHBs that coalesce by $z=0$.
\item Of these, the number of clusters in which BHBs are ejected before merger is approximately three times the number of clusters in which BHBs merge in-situ. 
\item Ex-situ BHB mergers originate from clusters of masses $10^5 \Msun - 10^{5.5}\Msun$. In-situ mergers occur in clusters over a much wider range of mass, with a peak at $10^{5.7} \Msun$.
\item The distribution of formation times for clusters that make merging BHBs peaks at $t=1-2$ Gyr, with a long tail extending towards later formation times; their metallicities span a wide range with a roughly flat distribution between $-2 < \logz < -0.5$. These distributions are similar to those of the overall cluster population.
\item 40\% of the clusters that produce merging BHBs will also disrupt by $z=0$. Most clusters are disrupted due to two-body relaxation driven evaporation, while a much smaller fraction inspiral into the NSC.
\item Using the mean properties for each cluster, and normalizing the number of BHBs per cluster at $\Mcl = 10^{5.5} \Msun$, the $z=0$ cosmic merger rate of dynamically assembled BHs is $\sim$ 6 Gpc$^{-3}$ yr$^{-1}$. 
\item The merger rate of dynamically assembled BHBs is weakly increasing out to $z\sim 1.5$ and drops at higher redshift. This behaviour is driven by dynamical processes within the cluster, which introduce a significant delay between cluster formation and BHB mergers.
\end{enumerate}
\begin{acknowledgements}
We thank Carl Rodriguez and the anonymous referee for comments and discussions that improved this work as well as Mario Spera \& Michaela Mapelli for making the \textsc{sevn} code publicly available. NC acknowledges support from the Centre for Cosmological studies and thanks Joe Silk \& Peter Behroozi for helping make this project possible as well as Goni Halevi for support throughout the preparation of this work. MV acknowledges funding from the European Research Council under the European Community's Seventh Framework Programme (FP7/2007-2013 Grant Agreement no.\ 614199, project ``BLACK''). OG was supported in part by the NSF through grant 1412144.
\end{acknowledgements}

\bibliographystyle{apj}
\bibliography{gc_nick} 

\end{document}